\documentclass[journal]{IEEEtran}
\usepackage{graphicx,cite,amssymb,amsmath,psfrag,subfigure}
\usepackage{graphicx}
\usepackage{amssymb}
\usepackage{amsmath,empheq}
\usepackage{cite}
\usepackage{subfigure}
\usepackage{mathrsfs}
\usepackage[displaymath,mathlines]{lineno}
\usepackage{color}
\usepackage{tabulary}
\usepackage{multirow}
\usepackage{cases}
\usepackage{stfloats}
\usepackage{url}

\usepackage{enumerate,color,graphicx,epsfig,subfigure,amsmath,amssymb}
\usepackage{mathtools}
\usepackage[mathcal]{euscript}

\newtheorem{proposition}{Proposition}

\newtheorem{remark}{Remark}
\newtheorem{algorithm}{Algorithm}

\DeclareMathOperator{\E}{\mathbb{E}}

\newcommand{\Ptotal}{P_{\mathrm{total}}}

\newcommand{\EX}[1]{\E\left\{{#1}\right\}}

\newcommand{\CG}[2]{\mathcal{CN}\left({#1},{#2}\right)}

\newcommand{\B}[1]{{\mathbf{#1}}}

\newcommand{\Pp}{\rho_{\mathrm{p}}}

\newcommand{\Pd}{\rho_{\mathrm{d}}}
\newcommand{\Se}{{\mathsf{ S}}_{\mathrm{e}}}
\newcommand{\Ee}{{\mathsf{ E}}_{\mathrm{e}}}
\newcommand{\So}{{\mathsf{ S}}_{\mathrm{o}}}
\newcommand{\barSo}{{ \overline{\mathsf{S}} }_{\mathrm{o}}}

\newcommand{\p}{\mathrm{p}}

\newcommand{\tauc}{\tau_\mathrm{c}}
\newcommand{\taup}{\tau_\mathrm{p}}

\newcommand{\U}{\mathcal{U}}
\newcommand{\A}{\mathcal{A}}

\makeatletter
\def\@setsize#1#2#3#4{
    \@nomath#1
    \let\@currsize#1
    \baselineskip #2
    \baselineskip \baselinestretch\baselineskip
    \parskip \baselinestretch\parskip
    \setbox\strutbox \hbox{
        \vrule height.7\baselineskip
            depth.3\baselineskip
            width\z@}
    \skip\footins \baselinestretch\skip\footins
    \normalbaselineskip\baselineskip#3#4}
\makeatother

\makeatletter
\newcommand{\setstretch}[1]{
    \def\baselinestretch{#1}%
    \@currsize
    }
\makeatother

\newcounter{eqncnt}

\newcounter{eqnback}
\begin{document}

\title{
    On the Total Energy Efficiency of Cell-Free Massive MIMO}
\author{ {Hien Quoc Ngo, \emph{Member, IEEE}, Le-Nam Tran, \emph{Senior Member, IEEE}, Trung Q. Duong, \emph{Senior Member, IEEE}, Michail Matthaiou, \emph{Senior Member, IEEE}, and Erik G. Larsson, \emph{Fellow, IEEE}}

\thanks{Manuscript received February 16, 2017; revised September 14, 2017; accepted
October 25, 2017. The associate editor coordinating the review of
this paper and approving it for publication was Dr.~Mohamad Awad.
The work of T.~Q.\ Duong was supported in part by the U.K. Royal Academy of Engineering Research Fellowship under Grant RF1415$\setminus$14$\setminus$22 and by the U.K. Engineering and Physical Sciences Research Council (EPSRC) under Grant EP/P019374/1. The work of M.~Matthaiou  was supported in part by the EPSRC under grant EP/P000673/1. The work of E.~G.\ Larsson was supported in part by the Swedish Research Council (VR) and ELLIIT. Part of this work was presented at the IEEE International
Workshop on Signal Processing Advances in Wireless Communications
(SPAWC) 2017 \cite{NTDML:17:SPAWC}.}

\thanks{
H.~Q.\ Ngo, T. Q. Duong, and M. Matthaiou  are with the Institute of Electronics, Communications and Information Technology (ECIT),  Queen's  University  Belfast, Belfast, BT3 9DT, U.K.    
 (Email: hien.ngo@qub.ac.uk, trung.q.duong@qub.ac.uk, m.matthaiou@qub.ac.uk).
}
\thanks{
L. N. Tran is  with the School of Electrical and Electronic Engineering, University College Dublin, Ireland. (Email: nam.tran@ucd.ie).
}
\thanks{E.~G.\ Larsson is with the Department of Electrical Engineering (ISY), Link\"{o}ping University, 581 83 Link\"{o}ping, Sweden (Email: erik.g.larsson@liu.se).
}

\thanks{Digital Object Identifier xxx/xxx}

}

\markboth{IEEE Transactions on Green Communications and Networking, Vol. XX, No. X, XXX
2017}{}

\maketitle

\begin{abstract} 
We consider the cell-free massive multiple-input multiple-output (MIMO) downlink, where a very large number of distributed multiple-antenna access points (APs) serve many single-antenna users in the same time-frequency resource. A simple (distributed) conjugate beamforming scheme is applied at each AP via the use of local channel state information (CSI). This CSI is acquired through  time-division  duplex  operation  and  the  reception  of  uplink  training signals transmitted by the users. We derive a closed-form expression for the spectral efficiency taking into account the effects of channel estimation errors and power control. This closed-form result enables us to analyze the effects of backhaul power consumption, the number of APs, and the number of antennas per AP on the total energy efficiency,  as  well  as,  to  design  an optimal power allocation algorithm. The optimal power allocation algorithm aims at maximizing the total energy efficiency, subject to a per-user spectral efficiency constraint and a per-AP power constraint. Compared with the equal power control, our proposed power allocation scheme can double the total energy efficiency. Furthermore, we propose AP selections schemes, in which each user chooses a subset of APs, to reduce the power consumption caused by the backhaul links. With our proposed AP selection schemes, the total energy efficiency increases significantly, especially for large numbers of APs. Moreover, under a requirement of good quality-of-service for all users, cell-free massive MIMO outperforms the colocated counterpart in terms of energy efficiency.
\end{abstract}


\section{Introduction} \label{Sec:Introduction}

The performance of  cellular networks is   typically limited by inter-cell interference. In particular,  users close to the cell boundaries suffer from strong  interference (relative to their desired signal power). Network multiple-input multiple-output (MIMO) (also referred to as distributed MIMO, coordinated multi-point transmission, and distributed antenna systems) can reduce such inter-cell  interference through   coherent cooperation between   base stations \cite{KFV:06:WCM}. In network MIMO, the base stations cooperate via advanced backhaul links to jointly transmit signals in the downlink and jointly detect signals in the uplink. However, it was shown in \cite{LHA:13:IT} that base station cooperation has fundamental limitations, i.e., even with full cooperation between the base stations, the spectral efficiency is
upper-bounded by a finite constant   when  the transmit power goes to infinity. However, this does not mean that cooperation has no benefits. Cooperation can still yield significant higher spectral efficiencies  and coverage probabilities compared to the case where   interference is  ignored \cite{LHA:13:IT}.
 Consequently, there has been a great deal of interest in network MIMO over the past decade \cite{BZGO:10:SP,KG:14:IT,GZZ:16:COM}.  The main challenges in the implementation of  network MIMO are the need for a substantial backhaul overhead,  high deployment  costs, and a
sufficiently capable central processing unit.
In particular,   precoded signals and channel state information (CSI) need to be shared among the base stations.

Recently,  cell-free massive MIMO has been  introduced as a practical and useful embodiment of the network MIMO concept \cite{NAYLM:16:WCOM}.
In cell-free massive MIMO, a large number of access points (APs) equipped with single or multiple antennas, and distributed over a large area, coherently serving a large number of users in the same time-frequency resource.
 As in (cellular) colocated  massive MIMO \cite{MLYN:16:Book}, cell-free massive MIMO exploits the favorable propagation and channel hardening properties when the number of APs is large to multiplex many users in the same time-frequency resource with small inter-user interference. Thus, it can offer a huge spectral efficiency with simple signal  processing. More importantly, in a cell-free massive MIMO configuration, the service antennas are brought close to the users, which yields a  high degree of macro-diversity and low path losses \cite{NAYLM:16:WCOM}; hence, many users   can be served simultaneously with  uniformly good quality-of-service.   Furthermore, it was shown in \cite{NAYLM:16:WCOM} that cell-free massive MIMO  has  significantly  better performance  than   conventional small-cell systems, where each user is served by a single  AP.  For these reasons, cell-free massive MIMO is  as a promising technology for next generation wireless systems. 

Despite its potential, however, besides \cite{NAYLM:16:WCOM} there is fairly little work on
cell-free massive MIMO available in the literature. In \cite{NAMHR:17:WCOM}, the performance of cell-free massive MIMO with zero-forcing processing was analyzed, under the assumption that  all pilot sequences assigned to the users were mutually orthogonal. Cell-free massive MIMO with beamformed downlink training  was investigated in \cite{INLF:16:GLOBECOM}. The conclusion was that by beamforming the pilots, the performance of the cell-free
massive MIMO downlink can be substantially improved. A compute-and-forward approach  for cell-free massive MIMO to reduce the backhaul load was proposed and analyzed in \cite{HB:16:ICC}. All the above-cited works assumed that the APs have only a single antenna. However,  the APs can be equipped with multiple antennas to increase the diversity and array gains, as well as,   reduce the backhaul requirements. In addition, while it is well-known that colocated massive MIMO is energy-efficient \cite{BSHD:15:WCOM},  it is not yet clear how good the energy efficiency of cell-free massive MIMO is. The argument is that in cell-free massive MIMO, more backhaul links are required  which  potentially increase the total power consumption to such a level that can overwhelm the spectral efficiency gains.  There are many works on the energy efficiency of network MIMO in the literature, such as \cite{XCZLY:15:JSAC,BSK:16:JSAC,RLPH:17:VT}. But to the authors' best knowledge, the total energy efficiency, that takes into account the effects of channel estimation, power control, AP selection schemes, hardware and backhaul power consumption, has not been previously exploited for cell-free massive MIMO.

Motivated by the above discussion, in this work we consider a cell-free massive MIMO system with multiple antennas at each AP and time-division duplex (TDD) operation. All APs cooperate via a backhaul network to jointly transmit signals to all users in the same time-frequency resource. The transmission is performed through the use of conjugate beamforming. The total energy efficiency of this system is investigated. The specific contributions of the paper are as follows:
\begin{itemize}
\item We derive a closed-form expression for the spectral efficiency of the downlink channel with any finite numbers of APs and users and arbitrary pilot sequences assigned to the users. Our result is a non-trivial generalization of the results for colocated massive MIMO in \cite{MLYN:16:Book} and cell-free massive MIMO with single-antenna APs in \cite{NAYLM:16:WCOM}.

\item We investigate the total energy efficiency of our considered system taking into account the hardware power consumption and the power consumption of the backhaul. We propose an optimal power control algorithm which aims at maximizing the
total energy efficiency, under a per-user spectral efficiency constraint and a per-AP power constraint. The solution to this power control problem can be approximately determined
by solving a sequence of second-order cone programs (SOCPs) \cite{Alizadeh2003}.

\item We study the effect of the backhaul power consumption, and show that the backhaul power consumption significantly affects the energy efficiency, especially when the number of APs is large. To reduce this effect, we propose two AP selection schemes: received-power-based selection and largest-large-scale-fading-based selection. Then, the performances of cell-free massive MIMO and colocated massive MIMO are quantitatively compared. 
\end{itemize}

The remainder of paper is organized as follows. In
Section~\ref{Sec:SysModel}, we describe the cell-free massive MIMO
system model and derive the downlink spectral efficiency. In Section~\ref{sec:eeff}, we present the power consumption model and the corresponding energy efficiency. The power control
and access point selection schemes are discussed in Sections~\ref{sec:totalEE} and  \ref{sec:APsel}, respectively.  Numerical results and discussions are provided in Section~\ref{sec:numerical_results}, while Section~\ref{sec:conclusion} concludes the paper. Table~\ref{table:notation} tabulates the notation and symbols used throughout the paper.


\begin{table}[t!]
    \caption{
      Notation and Symbols
    }
    \centerline{
\begin{tabular}{|l|l|}
\hline
$(\cdot)^T$, $(\cdot)^\ast$,$(\cdot)^H$  & Transpose, complex conjugate, transpose conjugate \\
$\|\cdot\|$ & Euclidean norm\\
$\mathbb{E}\left\{\cdot\right\}$ & Expectation operator\\
$[\B{x}]_n$ & The $n$-th element of vector $\B{x}$\\
$ \CG{0}{\sigma^2}$ &  Circularly
symmetric complex Gaussian distribution \\
 &  with zero
mean and variance $\sigma^2$\\
$\mathcal{N}(0,\sigma^2)$ &  Real-valued Gaussian distribution\\&  with zero
mean and variance $\sigma^2$\\
\hline
\end{tabular}}
    \label{table:notation}
\end{table}

\section{System Model and Spectral Efficiency} \label{Sec:SysModel}
\subsection{System Model}

\begin{figure}[t!]
        \begin{center}
        \begin{psfrags}
        \psfrag{AP1}[tc][Bl][1.0]{AP $1$}
        \psfrag{AP2}[tc][Bl][1.0]{AP $2$}
        \psfrag{APm}[tc][Bl][1.0]{AP $m$}
        \psfrag{APM}[tc][Bl][1.0]{AP $M$}
        \psfrag{CPU}[tc][Bl][1.0]{CPU}
        \psfrag{terminal 1}[tc][Bl][1.0]{user $1$}
        \psfrag{terminal k}[tc][Bl][1.0]{user $k$}
        \psfrag{terminal K}[tc][Bl][1.0]{user $K$}
        \psfrag{gmk}[Bl][Bl][1.0]{$g_{mk}$}
        \includegraphics[width=0.450\textwidth]{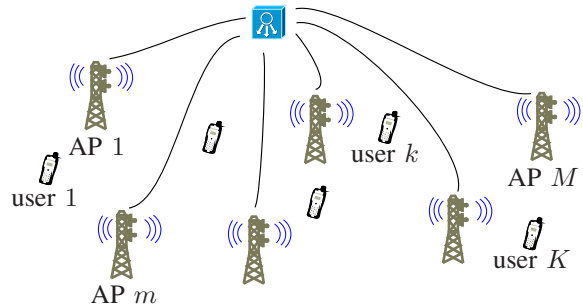}
        \end{psfrags}
\caption{Cell-free massive MIMO system.\label{fig:system}}
        \end{center}
\end{figure}

We consider a cell-free massive MIMO system (see Figure~\ref{fig:system}) where $M$ APs serve $K$ users in the same time-frequency resource under TDD operation. Each AP is equipped with $N$ antennas, while each user has a single antenna. We assume that the APs and the users are randomly located in a large area, and all APs are connected to a central processing unit (CPU) through a backhaul network. We further assume that $M\gg K$. With TDD CSI acquisition, the channel estimation overhead scales as $K$, and is independent of $M$. Therefore, $M$ can be made as large as desired, while $K$ is limited by mobility. (It is approximately
upper bounded by the coherence time divided by the channel
delay-spread.) As a result, operating with $M\gg K$ is both a desirable and a natural operating point.

For TDD operation in the context of massive MIMO, the existing literature  typically considers three phases within each coherence interval: uplink training, uplink payload data transmission, and downlink payload data transmission. In this work, we focus on the downlink, and thus the uplink payload data transmission phase is neglected. The system model is similar to that in \cite{NAYLM:16:WCOM}, but an important distinction is that the APs have multiple antennas herein. Let $\tauc$ be the length of each coherence interval (in samples). A part of the coherence interval is used for uplink training. The length of this duration is denoted by $\taup$, $\taup<\tauc$. The remaining duration, ($\tauc-\taup$), is used for the downlink payload data transmission.

\subsubsection{Uplink Training} \label{sec CF CEst}
All $K$ users simultaneously transmit their pilot sequences to all APs. Let $\sqrt{\taup}\pmb{\varphi}_k \in
\mathbb{C}^{\taup\times 1}$, where $\|\pmb{\varphi}_k\|^2=1$, be
the pilot sequence transmitted from the $k$-th user, $k=1,  \ldots, K$. Then, the $m$-th AP receives 
\begin{align}\label{eq:pilot1}
	\B{Y}_{\mathrm{p},m}  
	= 
	\sqrt{\taup
	\Pp}\sum_{k=1}^K \B{g}_{mk} \pmb{\varphi}_k^H +
	\B{W}_{\p,m},
\end{align}
where $\Pp$ is the normalized transmit signal-to-noise ratio
(SNR) of each pilot symbol, $\B{W}_{\p,m}$ is an $N\times \taup$ noise matrix whose elements are
independent and identically distributed (i.i.d.) $\CG{0}{1}$ RVs, and $\B{g}_{mk}\in\mathbb{C}^{N\times 1}$ is the channel vector between the $m$-th AP and the $k$-th user. The channel $\B{g}_{mk}\in\mathbb{C}^{N\times 1}$ models the propagation as follows:
\begin{align}\label{eq:gmk}
\B{g}_{mk} = \beta_{mk}^{1/2}\B{h}_{mk},
\end{align}
where $\beta_{mk}$ represents the large-scale fading which does not depend on the antenna indexes at the AP, and $\B{h}_{mk}$ is an $N\times 1$ vector of small-scale fading coefficients between the $N$ antennas of the $m$-th AP and the $k$-th user. We assume that small-scale fading is Rayleigh fading, i.e., the elements of $\B{h}_{mk}$ are i.i.d.\ $\CG{0}{1}$ RVs.

The minimum mean-square error  (MMSE) channel estimate of $\hat{\B{g}}_{mk}$,  given $\check{\B{y}}_{\p,mk}$, is \cite{KAY:93:Book}
\begin{align}\label{eq:MMSE est1}
\hat{\B{g}}_{mk} 
	&=
	\EX{\B{g}_{mk} \check{\B{y}}_{\p,mk}^H}  \left(\EX{ \check{\B{y}}_{\p,mk} \check{\B{y}}_{\p,mk}^H} \right)^{-1}
\check{\B{y}}_{\p,mk}\nonumber\\
	&= 
	\frac{\sqrt{\taup\Pp}\beta_{mk}}{\taup\Pp\sum_{k'=1}^K\beta_{mk'}\left|\pmb{\varphi}_{k'}^H
\pmb{\varphi}_{k}\right|^2+1}\check{\B{y}}_{\p,mk},
\end{align}
where\begin{align}\label{eq:tidley1}
\check{\B{y}}_{\mathrm{p},mk} 
	&\triangleq
\B{Y}_{\mathrm{p},m} \pmb{\varphi}_k \nonumber\\
	&= \sqrt{\taup
	\Pp} \B{g}_{mk}  + \sqrt{\taup
	\Pp}\sum_{k'\neq k}^K \B{g}_{mk'} \pmb{\varphi}_{k'}^H \pmb{\varphi}_k +
	\tilde{\B{w}}_{\p,mk},
\end{align}
where $\tilde{\B{w}}_{\p,mk}\triangleq \B{W}_{\p,m}\pmb{\varphi}_k$ includes i.i.d.\ $\CG{0}{1}$ components.

 The channel estimate $\hat{\B{g}}_{mk}$ includes $N$ independent and identical Gaussian components. The mean-square of the $n$-th component is denoted by $\gamma_{mk}$, and given by
\begin{align}\label{eq:gamma1}
\gamma_{mk}
	&\triangleq \EX{\left|\left[\hat{\B{g}}_{mk}\right]_n \right|^2}\nonumber\\
 &=
\frac{{\taup\Pp}\beta_{mk}^2}{\taup\Pp\sum_{k'=1}^K\beta_{mk'}\left|\pmb{\varphi}_{k'}^H
\pmb{\varphi}_{k}\right|^2+1}.
\end{align}
Denote by $\tilde{\B{g}}_{mk} ={\B{g}}_{mk} -\hat{\B{g}}_{mk}$ the channel estimation error. From the MMSE estimation property, $\tilde{\B{g}}_{mk}$ is independent of $\hat{\B{g}}_{mk}$. The elements of $\tilde{\B{g}}_{mk}$ are i.i.d.\ $\CG{0}{\beta_{mk}-\gamma_{mk}}$ RVs.

\subsubsection{Downlink Payload Data Transmission}

After acquiring the channels from the uplink pilots, the APs use
conjugate beamforming to transmit signals to the $K$ users. Our choice of conjugate beamforming is inspired by the fact that it is computationally  simple with most processing done locally at the APs. More precisely,
there is no need for exchanging the instantaneous CSI among the APs or the central unit. Furthermore, when the number of APs is large, conjugate beamforming offers excellent performance [6].

Denote the symbol intended for the $k$-th user by $q_k$, where $\EX{|q_k|^2}=1$, $k=1, \ldots, K$. The vector of transmitted signals from the $m$-th AP, $\B{x}_m$, is generated by first scaling the $K$ symbols with the power control coefficients $\{\eta_{mk}\}$, and then multiplying them with the conjugate of the channel estimates as follows:
\begin{align}\label{eq:xm}
\B{x}_m = \sqrt{\Pd}\sum_{k=1}^K
\sqrt{\eta_{mk}}\hat{\B{g}}_{mk}^\ast q_k,
\end{align}
where $\Pd$ is the maximum normalized transmit power (normalized by the noise power $N_0$) at each AP.
The normalized transmitted power  is 
\begin{align}\label{eq:xm1}
\EX{\|\B{x}_m\|^2} = \Pd N\sum_{k=1}^K
\eta_{mk}\gamma_{mk}.
\end{align}
The power control coefficients $\{\eta_{mk}\}$ are chosen to satisfy the power constraint at each AP,
$
\EX{\|\B{x}_m\|^2}\leq \Pd.
$
Thus,
\begin{align}\label{eq:pct}
\sum_{k=1}^K \eta_{mk}\gamma_{mk} \leq \frac{1}{N}, \quad \text{for all
$m = 1, \ldots, M$}.
\end{align}
With the transmitted signal vector $\B{x}_m$ given in \eqref{eq:xm}, the $k$-th user receives
\begin{align}\label{eq:rk1}
r_{k} 
	&= 
	\sum_{m=1}^M \B{g}_{mk}^T \B{x}_m +  w_{k} \nonumber\\
	&=
	\sqrt{\Pd}\sum_{m=1}^M
\sqrt{\eta_{mk}}\B{g}_{mk}^T\hat{\B{g}}_{mk}^\ast q_{k}\nonumber\\
	&+
\sqrt{\Pd}\sum_{k'\neq k}^K\sum_{m=1}^M
\sqrt{\eta_{mk'}}\B{g}_{mk}^T\hat{\B{g}}_{mk'}^\ast q_{k'} + w_{k},
\end{align}
where $w_{k} \sim \CG{0}{1}$ is the additive noise. 

\subsection{Spectral Efficiency}
The $k$-th user will detect its desired signal $q_k$ from the received signal $r_{k}$ given by \eqref{eq:rk1}. To do so, the $k$-th user needs to know the effective channel gain $\sum_{m=1}^M
\sqrt{\eta_{mk}}\B{g}_{mk}^T\hat{\B{g}}_{mk}^\ast$. Since there are no downlink pilots, the $k$-th user uses its knowledge of the channel statistics to detect $q_k$. More precisely, the $k$-th user treats the mean of the effect channel gain as the true channel for signal detection.  The benefits of relying on the channel statistics only are: (i) the need for downlink training is avoided; and (ii) a
simple closed-form expression for the spectral efficiency can be
derived which enables us to obtain important insights and to further
design power control, pilot assignment, and user scheduling
algorithms. Furthermore, since the number of APs is large, the
effective channel gain fluctuates only slightly around its mean (this
is a consequence of the law of large numbers). Consequently, detection
using only the channel statistics performs well \cite[Remark~4]{NAYLM:16:WCOM}. Note that this technique has been widely used in the massive MIMO context \cite{MLYN:16:Book}. With this technique, the received signal $r_k$ given in \eqref{eq:rk1} can be rewritten as
\begin{align}\label{eq:rk1r}
&r_{k} 
	=
	\sqrt{\Pd}\EX{\sum_{m=1}^M
\sqrt{\eta_{mk}}\B{g}_{mk}^T\hat{\B{g}}_{mk}^\ast} q_{k}\nonumber\\
	&\!+\!
	\sqrt{\Pd}\!\left(\!\sum_{m=1}^M
  \sqrt{\eta_{mk}}\B{g}_{mk}^T\hat{\B{g}}_{mk}^\ast\!-\!
  \EX{\sum_{m=1}^M  \sqrt{\eta_{mk}}\B{g}_{mk}^T\hat{\B{g}}_{mk}^\ast}\!\right) q_k\nonumber\\
	&+
\sqrt{\Pd}\sum_{k'\neq k}^K\sum_{m=1}^M
\sqrt{\eta_{mk'}}\B{g}_{mk}^T\hat{\B{g}}_{mk'}^\ast q_{k'} \!+\! w_{k}.
\end{align}

By using the capacity bound in \cite[Section~2.3]{MLYN:16:Book}, the corresponding spectral efficiency (expressed in bit/s/Hz) of the $k$-th user is given by
\begin{align}\label{eq:rateexpr1}
{\Se}_{k}
 \!=\!
 \frac{\tauc\!-\!\taup}{\tauc}\log_2
    \!\!\left(\!
    1 \!+\! \frac{\left|{\tt DS}_k \right|^2}{\EX{|{\tt BU}_k|^2} + \sum\limits_{k'\neq k}^K  \EX{|{\tt UI}_{kk'}|^2} + 1}
    \!\right),
\end{align}
where ${\tt DS}_k$, ${\tt BU}_k$, and  ${\tt UI}_{kk'}$ represent the desired signal, the beamforming uncertainty gain, and the inter-user interference, respectively, given by 
\begin{align}\label{eq:rat2a}
    {\tt DS}_k &\triangleq  \sqrt{\Pd}\EX{\sum_{m=1}^M \sqrt{\eta_{mk}}\B{g}_{mk}^T\hat{\B{g}}_{mk}^\ast},\\\label{eq:rat2b}
    {\tt BU}_k &\triangleq \sqrt{\Pd}\left(\sum_{m=1}^M
  \sqrt{\eta_{mk}}\B{g}_{mk}^T\hat{\B{g}}_{mk}^\ast \right.\nonumber\\&\left.-
  \EX{\sum_{m=1}^M  \sqrt{\eta_{mk}}\B{g}_{mk}^T\hat{\B{g}}_{mk}^\ast}\right),\\\label{eq:rat2c}
    {\tt UI}_{kk'} &\triangleq \sqrt{\Pd}\sum_{m=1}^M   \sqrt{\eta_{mk'}}\B{g}_{mk}^T\hat{\B{g}}_{mk'}^\ast.
\end{align}

In the following, we present an exact closed-form expression for the spectral efficiency \eqref{eq:rateexpr1}.

\begin{proposition}\label{prop1}
The spectral efficiency of the  transmission  from  the  APs  to  the $k$-th  user given in \eqref{eq:rateexpr1} can be represented in closed-form  as in \eqref{eq:Theo_rateexpr1} shown at the top of the page, where $\bar{\pmb{\eta}}_k \triangleq  [\sqrt{{\eta}_{1k}},\ldots,\sqrt{{\eta}_{Mk}}]^{T}\in \mathbb{R}^{M}_{+} $, consists of all power control coefficients associated with  user $k$, $\pmb{D}_{k'k}$ is a $\mathbb{R}^{M\times M}$ diagonal
	matrix whose $m$-th diagonal entry is given by $[\mathbf{D}_{k'k}]_{m,m}=\sqrt{\gamma_{mk'}\beta_{mk}}$, and
$$\bar{\pmb{\gamma}}_{k'k}\triangleq |\pmb{\varphi}_{k'}^{H}\pmb{\varphi}_{k}|\left[\gamma_{1k'}\frac{\beta_{1k}}{\beta_{1k'}},\gamma_{2k'}\frac{\beta_{2k}}{\beta_{2k'}},\ldots,\gamma_{Mk'}\frac{\beta_{Mk}}{\beta_{Mk'}}\right]^{T}.$$
\begin{IEEEproof}
See Appendix~\ref{app:rate}. 
\end{IEEEproof}
\end{proposition}

\setcounter{eqnback}{\value{equation}} \setcounter{equation}{14}
\begin{figure*}[!t]

\begin{align}\label{eq:Theo_rateexpr1}
{\Se}_{k}\left(\{\eta_{mk}\}\right)
 =
 \frac{\tauc-\taup}{\tauc}
\log_2\left(
	1+\frac{\Pd N^2|\bar{\pmb{\gamma}}_{kk}^{T}\bar{\pmb{\eta}}_{k}|^{2}}{\Pd N^2\sum\limits _{k'\neq k}^{K}\!|\bar{\pmb{\gamma}}_{k'k}^{T}\bar{\pmb{\eta}}_{k'}|^{2}+\Pd N\sum\limits _{k'=1}^{K}||\pmb{D}_{k'k}\bar{\pmb{\eta}}_{k'}||^{2}_{2}+1}
	\right),
\end{align}
\hrulefill
\end{figure*}
\setcounter{eqncnt}{\value{equation}}
\setcounter{equation}{\value{eqnback}}

Note that in the special case that all APs are equipped with a single antenna ($N=1$), the spectral efficiency \eqref{eq:Theo_rateexpr1}    is identical to the spectral efficiency in \cite[Eq.~(24)]{NAYLM:16:WCOM}. If we replace $N$ with $M$ and $M$ with $1$, we have the spectral efficiency for the colocated massive MIMO system in \cite[Table~3.2]{MLYN:16:Book}.

\begin{remark}
If the channel coherence interval is long enough (which corresponds to low mobility environments), then we can choose $\taup\geq K$ so that all $K$ pilot sequences $\pmb{\varphi}_{1}, \ldots, \pmb{\varphi}_{K}$ are pairwisely orthogonal. In this case, the pilot contamination term $\Pd N^2\sum\limits _{k'\neq k}^{K}\!|\bar{\pmb{\gamma}}_{k'k}^{T}\bar{\pmb{\eta}}_{k'}|^{2}$ in \eqref{eq:Theo_rateexpr1} disappears, and hence, the spectral efficiency can increase without bound when the number of APs increases. However, typically, the channel coherence interval is not long enough to allow for orthogonality among the $K$ pilot sequences. As a result, the spectral efficiency is bounded even when $M\to\infty$ (since the term $\Pd N^2\sum\limits _{k'\neq k}^{K}\!|\bar{\pmb{\gamma}}_{k'k}^{T}\bar{\pmb{\eta}}_{k'}|^{2}$ increases with the same rate as the desired signal power term $\Pd N^2|\bar{\pmb{\gamma}}_{kk}^{T}\bar{\pmb{\eta}}_{k}|^{2}$). This causes the so-called pilot contamination effect. 
\end{remark}

The sum spectral efficiency is given by
\setcounter{equation}{15}\begin{align}\label{eq:SE1}
    {\Se}\left(\{\eta_{mk}\}\right) = \sum_{k=1}^K
    {\Se}_{k}(\{\eta_{mk}\}).
\end{align}
\section{Power Consumption Model and Energy Efficiency}\label{sec:eeff}

\subsection{Power Consumption Model}
The total power consumption is modeled as \cite{TVZ:11:WCM,BSHD:15:WCOM,ZZYJL:16:VT,DY:16:JSAC}
\begin{align} \label{eq:totalP1}
\Ptotal =  \sum_{m=1}^M P_m + \sum_{m=1}^M P_{\text{bh},m},
\end{align}
where $P_m$ is the power consumption at the $m$-th AP due to the amplifier and the circuit 
power consumption part (including the  power  consumption  of  the  transceiver chains and the power consumed for signal processing), and $P_{\text{bh},m}$ is the power consumed by the backhaul link connecting the CPU  and the $m$-th AP. The power consumption $P_m$ can be modeled as
\begin{align} \label{eq:totalP2}
P_m = \frac{1}{\alpha_m} \Pd N_0  \left(N \sum_{k=1}^K \eta_{mk}\gamma_{mk}\right) + N P_{\text{tc},m},
\end{align}
where $0<\alpha_m \leq 1$ is the power amplifier efficiency, $N_0$ is the noise power, and $P_{\text{tc},m}$ is the internal power required to run the circuit components (e.g. converters, mixers, and filters) related to each antenna of the $m$-th AP.

 The backhaul is used to transfer the data between the APs and the CPU, and its power consumption is proportional to the sum spectral efficiency,
\begin{align} \label{eq:totalP3}
P_{\text{bh},m} = P_{0,m}  +B \cdot \Se\left(\{\eta_{mk}\}\right) \cdot P_{\text{bt},m},
\end{align}
where $P_{0,m}$ is  a  fixed  power  consumption  of  each  backhaul (traffic-independent power) which may depend on the distances between the APs and the CPU and the system topology,  $P_{\text{bt},m}$
is  the  traffic-dependent  power  (in  Watt per bit/s), and $B$
is  the  system
bandwidth.\footnote{The backhaul is also used to transfer the power allocation coefficients, synchronization signals, etc. This is done once per large-scale fading realization which stays constant for many coherence intervals. Therefore, we neglect the power consumed by this  processing.}

The substitution of \eqref{eq:totalP2} and \eqref{eq:totalP3} into \eqref{eq:totalP1} yields
\begin{align} \label{eq:totalP4} 
&\Ptotal = \Pd N_0\sum_{m=1}^M \frac{1}{\alpha_m } \left(N\sum_{k=1}^K \eta_{mk}\gamma_{mk}\right) \nonumber\\&+ \sum_{m=1}^M \!\!\left(N P_{\text{tc},m} \!+\!  P_{0,m}\right) + B \!\left(\sum_{m=1}^M P_{\text{bt},m}\right) \!{\Se\left(\{\eta_{mk}\}\right)}.
\end{align}

\subsection{Total Energy Efficiency}
The total energy efficiency (bit/Joule) is defined as the sum throughput (bit/s) divided by the total power consumption (Watt) in the network:
\begin{align}\label{eq:EE1}
    \Ee\left(\{\eta_{mk}\}\right) = \frac{{B\cdot\Se\left(\{\eta_{mk}\}\right)}}{ \Ptotal
    },
\end{align}
where $B$ is again the system bandwidth.

\section{Total Energy Efficiency Maximization}\label{sec:totalEE}

We aim at allocating the power coefficients $\{\eta_{mk}\}$ to maximize the total energy efficiency, under the
constraints on per-user spectral efficiency and transmit power at
each AP. More precisely, the optimization problem is formulated
as follows:
\begin{align} \label{eq:opt1}
   ( \mathcal{P}) : \left\{%
\begin{array}{ll}
  \mathop {\max}\limits_{\{\eta_{mk}\}} & {\Ee}(\{\eta_{mk}\})\\
	  \text{s.t.} & {\Se}_{k}(\{\eta_{mk}\})\geq {\So}_{k},~ \forall k,\\
              & \sum_{k=1}^K \eta_{mk}\gamma_{mk} \leq 1/N, ~ \forall m, \\
              & \eta_{mk} \geq 0, ~ \forall k, ~ \forall m,\\
\end{array}%
\right.
\end{align}
where ${\So}_{k}$ is the minimum spectral efficiency required by the $k$-th user. 

Denote by $\bar{P}_{\mathrm{fix}} \triangleq  \sum\limits_{m=1}^M \left(N P_{\text{tc},m} +  P_{0,m}\right)$. Then, by following Appendix~\ref{app_trans1}, the optimization problem $(\mathcal{P})$ is equivalent to 
\begin{subequations} \label{eq:opt1cd}
\begin{empheq}[left={({\mathcal{P}_1}) :}\empheqlbrace]{align}%
  \mathop {\max}\limits_{\{\eta_{mk}\}} & \quad \frac{{B\cdot\Se(\{\eta_{mk}\})}}{\bar{P}_{\mathrm{fix}} + \Pd N_0 N\sum\limits_{m=1}^M \frac{1}{\alpha_m } \sum\limits_{k=1}^K \eta_{mk}\gamma_{mk} 
    } \label{eq:obj}\\
  \text{s.t.} & \quad {\Se}_{k}(\{\eta_{mk}\})\geq {\So}_{k},~ \forall k, \label{eq:QoS1}\\
  &\quad \sum_{k=1}^K \eta_{mk}\gamma_{mk} \leq 1/N, ~\forall m, \label{eq:maxpower1}\\
  &\quad \eta_{mk} \geq 0, ~ \forall k, ~ \forall m.\label{eq:power1}
  \end{empheq}
\end{subequations}    
    
 We remark  that the problem $(\mathcal{P}_1)$ is nonconvex since ${\Se}_{k}(\{\eta_{mk}\})$ is neither convex nor concave with respect to $\{\eta_{mk}\}$. Thus,  sequential convex approximation (SCA) will be applied to find a high-performance solution. To arrive at a more tractable formulation, we now show that  \eqref{eq:QoS1} indeed admits an equivalent convex expression. Towards this end, denote by $c_{mk}\triangleq \sqrt{\eta_{mk}}$, $\B{c}_k \triangleq  [c_{1k},\ldots,c_{Mk}]^{T}$, and $\B{c}\triangleq [\B{c}_1^T, \ldots, \B{c}_K^T]^T$. Then, the optimization problem $(\mathcal{P}_1)$ can be rewritten as
\begin{subequations} \label{eq:opt1c}
\begin{empheq}[left={({\mathcal{P}_1}) :}\empheqlbrace]{align}%
  \mathop {\max}\limits_{\B{c}} & \quad \frac{{B\cdot\Se(\B{c})}}{\bar{P}_{\mathrm{fix}} + \Pd N_0 N\sum\limits_{m=1}^M \frac{1}{\alpha_m } \sum\limits_{k=1}^K c_{mk}^2\gamma_{mk} 
    } \label{eq:obj2}\\
  \text{s.t.} & \quad {\Se}_{k}\left(\B{c}\right)\geq {\So}_{k},~ \forall k, \label{eq:QoS}\\
  &\quad \sum_{k=1}^K c_{mk}^2\gamma_{mk} \leq 1/N, ~ \forall m, \label{eq:maxpower}\\
  &\quad c_{mk} \geq 0, ~ \forall k, ~ \forall m,\label{eq:power}
  \end{empheq}
\end{subequations}     
    where ${\Se}_{k}\left(\B{c}\right)$ is given as  \eqref{eq:ratechange1}, shown at the top of the next page.

\setcounter{eqnback}{\value{equation}} \setcounter{equation}{24}
\begin{figure*}[!t]
\begin{align}\label{eq:ratechange1}
	{\Se}_{k}\left(\B{c}\right) 
	\triangleq 
	 \left(1-\frac{\taup}{\tauc}\right)
\log_2\left(
	1+\frac{\Pd N^2|\bar{\pmb{\gamma}}_{kk}^{T}\B{c}_{k}|^{2}}{\Pd N^2\sum\limits _{k'\neq k}^{K}\!|\bar{\pmb{\gamma}}_{k'k}^{T}\B{c}_{k'}|^{2}+\Pd N\sum\limits _{k'=1}^{K}||\pmb{D}_{k'k}\B{c}_{k'}||^{2}_{2}+1}
	\right).	
\end{align}
\hrulefill
\end{figure*}
\setcounter{eqncnt}{\value{equation}}
\setcounter{equation}{\value{eqnback}}

It is now clear  that \eqref{eq:QoS} is equivalent to the following second order cone (SOC) constraint 
\setcounter{equation}{25}\begin{align}\label{eq:QoS:SOC}
& \left|\bar{\pmb{\gamma}}_{kk}^{T}\B{c}_{k}\right|^2 
 \geq \nonumber\\
&\left(2^{{\barSo}_{k}}\!-\!1\right)\!\!\!\left( \sum\limits _{k'\neq k}^{K}\!|\bar{\pmb{\gamma}}_{k'k}^{T}\B{c}_{k'}|^{2}+\frac{1}{N}\!\!\sum\limits _{k'=1}^{K}||\pmb{D}_{k'k}\B{c}_{k'}||^{2}_{2}+\frac{1}{N^2\Pd}  \!\right),
\end{align}
where ${\barSo}_{k} \triangleq \frac{\tauc}{\tauc-\taup}{\So}_k$.

Next we further rewrite  $(\mathcal{P}_1)$ as the following  optimization problem
\begin{subequations} \label{eq:opt4}
\begin{empheq}[left={(\hat{\mathcal{P}}_1) :}\empheqlbrace]{align}%
\mathop {\max}\limits_{\B{c},\mathbf{t},t_0} & \quad B \frac{\sum_{k=1}^{K} t_k}{t_0} \label{eq:obj:subopt1}\\
\text{s.t.} & \quad  {\Se}_{k}\left(\B{c} \right)\geq t_k, ~ \forall k, \label{eq:SINR:epi:subopt1}\\
&\quad \bar{P}_{\mathrm{fix}} + \frac{\Pd N_0 N}{\alpha_m} \sum\limits_{m=1}^M   \sum\limits_{k=1}^K c_{mk}^2\gamma_{mk}  \leq t_{0} \label{eq:power:epi:subopt1}\\
&\quad \eqref{eq:maxpower}, \eqref{eq:power}, \eqref{eq:QoS:SOC},
\end{empheq}
\end{subequations}
where $\B{t} \triangleq [t_1, \ldots, t_K]^T$. It is easy to see that if $(c_{mk},\mathbf{t},t_0)$ solves $(\hat{\mathcal{P}}_1)$, then $c_{m,k}$ solves  $({\mathcal{P}}_1)$. The proof is due to the fact that at optimality, all the constraints \eqref{eq:SINR:epi:subopt1} and \eqref{eq:power:epi:subopt1} hold with equality. Our motivation for the above maneuver is twofold. First, the objective in \eqref{eq:opt4} is a linear fractional function which is much easier to handle, from a viewpoint of  Charnes-Cooper transformation that we shall show shortly. Second, the reformulation given in \eqref{eq:opt4}  facilitates a customization of the branch-and-bound method based on monotonic optimization to find an optimal solution, which is described next.

Specifically, the formulation of  $(\hat{\mathcal{P}}_1)$  reveals  three important observations:  (i) the objective in \eqref{eq:obj:subopt1} is monotonically increasing with $\mathbf{t}$ and $1/t_0$;   (ii) for a given fixed $\mathbf{t}$ and $1/t_0$, the constraints in \eqref{eq:SINR:epi:subopt1} and \eqref{eq:power:epi:subopt1}  are convex; and (iii) the constraints \eqref{eq:maxpower},  \eqref{eq:power}, \eqref{eq:QoS:SOC} are convex. These three facts simply mean that a globally optimal solution can be  found using a monotonic optimization method, i.e., the branch-and-reduce-and bound (BRB) method  \cite{TAT:05:bookchapter}. The description of a BRB for solving $(\hat{\mathcal{P}}_1)$ is quite involved and thus is omitted for the sake of brevity.  The interested reader is referred to  \cite{TAT:05:bookchapter,TTJ:16:SP} for further details.  However, such a global optimization method generally induces very high computational complexity, especially for our cell-free massive MIMO system where the number of APs and  the number of users are very large. Therefore, we propose a sub-optimal solution which has low computational complexity and is shown to achieve a performance close to the optimal one. The method is based on the SCA method \cite{BBT:10:J_GO,SFL:16:SP,NHTJ:12:SPL}.

It is obvious that the troublesome constraint in \eqref{eq:opt4} is \eqref{eq:SINR:epi:subopt1}, which  is non-convex. To deal with this nonconvex constraint, we introduce the slack variables $ u_k $, $ k=1,\ldots, K$, and rewrite \eqref{eq:SINR:epi:subopt1} as 
\begin{subequations} \begin{eqnarray}
&&\hspace{-0.5cm}1\!+\!\frac{ N^2 |\bar{\pmb{\gamma}}_{kk}^{T}\B{c}_{k}|^{2}}{ N^2\!\!\!\sum\limits _{k'\neq k}^{K}\!\!|\bar{\pmb{\gamma}}_{k'k}^{T}\B{c}_{k'}|^{2}+ N\!\!\! \sum\limits _{k'=1}^{K}\!\!\|\pmb{D}_{k'k}\B{c}_{k'}\|^{2}_{2}\!+\!\frac{1}{\Pd}} \geq  u_{k},\label{eq:SINR:slack}\\
&&\hspace{-0.5cm}\log_2( u_{k} ) \!\geq \! \frac{\tauc}{\tauc-\taup} t_{k}.\label{eq:rate:cpt}
\end{eqnarray}
\end{subequations}
The constraint in \eqref{eq:SINR:slack} is equivalent to 
 \begin{equation}\label{eq:SINR:slack:equi}
  f(\B{c},u_k) \geq  \Pd N^2\sum\limits _{k'\neq k}^{K}|\bar{\pmb{\gamma}}_{k'k}^{T}\B{c}_{k'}|^{2}+\Pd N \sum\limits _{k'=1}^{K}||\pmb{D}_{k'k}\B{c}_{k'}||^{2}_{2}+1,
 \end{equation}
 where 
 \begin{equation}\label{eq:quadoverlin}
 f(\B{c},u_k)
 \triangleq 
 \frac{\Pd N^2\!\sum\limits _{k'=1}^{K}|\bar{\pmb{\gamma}}_{k'k}^{T}\B{c}_{k'}|^{2}+\Pd N\!\sum\limits _{k'=1}^{K}||\pmb{D}_{k'k}\B{c}_{k'}||^{2}_{2}+1}{u_k}.
 \end{equation}
 Note that $ f(\B{c},u_k) $ defined in \eqref{eq:quadoverlin} is a quadratic-over-linear function which is jointly convex in $ \B{c}$ and $ u_k  $. In light of SCA, we can approximate it by a first-order Taylor expansion of \eqref{eq:quadoverlin}. Specifically, let  $\B{c}^{n}  $ and $ u_{k}^{n}  $ be the values of $\B{c}$ and $ u_k  $ after $ n $ iterations of the SCA process. Then, we can replace \eqref{eq:SINR:slack:equi} with 
 \begin{align}\label{eq:rate:1storder}
 F(\B{c},u_k;\B{c}^{n},u_k^{n}) &\geq  \Pd N^2 \sum\limits _{k'\neq k}^{K}|\bar{\pmb{\gamma}}_{k'k}^{T}\B{c}_{k'}|^{2}\nonumber\\&+\Pd N\sum\limits _{k'=1}^{K}||\pmb{D}_{k'k}\B{c}_{k'}||^{2}_{2}+1,
 \end{align}
 where
\begin{align}
 &F(\B{c},u_k;\B{c}^{n},u_k^{n})=
 f(\B{c}^{n},u_k^{n}) \nonumber\\
 &+ \nabla_{\B{c}} f(\B{c}^{n},u_k^{n})^{T} (\B{c}-\B{c}^{n})+\partial_{u_k} f(\B{c}^{n},u_k^{n}) (u_k-u_k^{n})\nonumber\\
 &=
 \frac{2\Pd}{u_k^n}\sum\limits _{k'=1}^{K}\B{c}^{nT}_{k'}\left( N^2\bar{\pmb{\gamma}}_{k'k}\bar{\pmb{\gamma}}_{k'k}^{T} + N\pmb{D}_{k'k}^2\right)\left(\B{c}_{k'}-\B{c}_{k'}^{n}\right)\nonumber\\
&+ f(\B{c}^{n},u_k^{n}) -
 \frac{f(\B{c}^{n},u_k^{n})}{u_k^{n}}(u_k-u_k^{n}),
\end{align}
where in the last equality we have used the identity 
$$\nabla_{\B{x}}\left(\B{x}^T\B{A}\B{x}\right) = \left(\B{A}+\B{A}^T\right)\B{x}.$$ Note that \eqref{eq:rate:1storder} is SOC representable \cite{BV:04:Book}. 

The constraint  \eqref{eq:rate:cpt} deserves special attention. In fact it is a convex constraint and thus convex approximation is not necessary  as convexity should be preserved. In this way, however, the resulting convex program cannot be cast into a more standard form for which powerful solvers are available. More specifically, if \eqref{eq:rate:cpt} is kept as it is, then we obtain a convex problem that is a mix of second order and exponential cones. Solvers for such a convex problem do exist \cite{DCBL:13:ECC}, but our numerical experiments reveal that they are not scalable with the problem size. To arrive at a more computationally efficient formulation, our idea  is to approximate \eqref{eq:rate:cpt} by a more tractable constraint, i.e., an SOC one. Accordingly to the SCA principle, we need to find a convex lower bound of the right hand side of \eqref{eq:rate:cpt}. To this end, we recall the following inequality    
 $$\ln(x) \geq 1 - \frac{1}{x},$$ 
which leads to
\begin{align}
	\log_2( u_{k} ) \geq  \log_2(u_k^n) + \log_2(e)\left(1 - \frac{u_k^n}{u_k}\right).\label{logbound}
\end{align}
Note that the above inequality holds with equality when $ {u_k}=u_k^n $. Moreover, the first derivative of both sides of \eqref{logbound} is the same  when $ {u_k}=u_k^n $. That is, the  right hand side of \eqref{logbound} is a proper convex bound in light of SCA \cite{BBT:10:J_GO,SFL:16:SP}.
Therefore, the constraint \eqref{eq:rate:cpt} can be replaced by
\begin{align}\label{eq:ueq1}
	 \log_2(u_k^n) + \log_2(e)\left(1 - \frac{u_k^n}{u_k}\right)
	 \geq \frac{\tauc}{\tauc-\taup} t_{k}.
\end{align}
We remark that the above constraint can be reformulated as an SOC constraint \cite{Alizadeh2003}.
 In summary, the problem at the $ (n+1) $-th iteration of the proposed method is given by
\begin{subequations} \label{eq:subprob}
\begin{empheq}[left={(\hat{\mathcal{P}}_{1,n+1}) :}\empheqlbrace]{align}%
\mathop {\max}\limits_{c_{m,k},\mathbf{t},\mathbf{u}} & \quad B \frac{\sum_{k=1}^{K} t_k}{t_0} \label{eq:obj:subopt}\\
\text{s.t.} & \quad \eqref{eq:maxpower},\eqref{eq:power},\\
 & \quad \eqref{eq:QoS:SOC},\eqref{eq:power:epi:subopt1},\eqref{eq:rate:1storder},\eqref{eq:ueq1}.
\end{empheq}
\end{subequations}

By using a perspective transformation, $ (\hat{\mathcal{P}}_{1,n+1}) $ can be reformulated as an SOCP as \eqref{eq:subprob:cvx}, shown at the top of the next page.

\setcounter{eqnback}{\value{equation}} \setcounter{equation}{35}
\begin{figure*}[!t]
\begin{subequations} \label{eq:subprob:cvx}
\begin{empheq}[left={(\hat{\mathcal{P}}_{1,n+1}) :}\empheqlbrace]{align}%
\mathop {\max}\limits_{\dot{c}_{m,k},\dot{\mathbf{t}},\dot{\mathbf{u}},\theta} &  \quad  B {\sum_{k=1}^{K} \dot{t}_{k}} \\
\text{s.t.} & \quad \sum_{k=1}^{K}\dot{c}_{mk}^{2}\gamma_{mk}  \leq \theta^2/N,~m=1,\ldots,M\\
&\quad \dot{c}_{mk} \geq  0,~k=1,\ldots,K,~M=1,\ldots,M,\\
&\hspace{-20pt} {|\bar{\pmb{\gamma}}_{kk}^{T}\dot{\B{c}}_{k}|}  \geq  \sqrt{(2^{{\barSo}_{k}}-1)\left(\sum\limits _{k'\neq k}^{K}\!|\bar{\pmb{\gamma}}_{k'k}^{T}\dot{\B{c}}_{k'}|^{2}+\frac{1}{N}\sum\limits _{k'=1}^{K}||\pmb{D}_{k'k}\dot{\B{c}}_{k'}||_{2}^{2}+\frac{\theta^{2}}{N^2\Pd}\right)}\\
&\bar{P}_{\mathrm{fix}}\theta^{2}+\Pd N_{0} N\sum\limits _{m=1}^{M}\frac{1}{\alpha_m}\sum\limits _{k=1}^{K}\dot{c}_{mk}^{2}\gamma_{mk} \leq  \theta\\
&	 \theta\left(\log_2(u_k^n) + \log_2(e)\right)
	 \geq \frac{\tauc}{\tauc-\taup} \dot{t}_{k} + \log_2(e)\frac{u_k^n}{\dot{u}_k}\theta^2, \quad k= 1, \ldots, K, \\
&\theta \bar{F}(\dot{\B{c}},\dot{u}_{k};\B{c}^{n},u_{k}^{n}) \geq  \Pd N^2\sum\limits _{k'\neq k}^{K}|\bar{\pmb{\gamma}}_{k'k}^{T}\dot{\B{c}}_{k'}|^{2}\!+\!\Pd N \sum\limits _{k'=1}^{K}||\pmb{D}_{k'k}\dot{\B{c}}_{k'}||^{2}_{2}+\theta^2, \label{eq:rotatecone}
\end{empheq}
\end{subequations}
where $\dot{\B{c}}_k \triangleq  [\dot{c}_{1k},\ldots,\dot{c}_{Mk}]^{T}$, $\dot{\B{c}}\triangleq [\dot{\B{c}}_1^T ~ \ldots ~ \dot{\B{c}}_K^T]^T$, and
\begin{align}\label{}
&\bar{F}(\dot{\pmb{{\eta}}},\dot{u}_{k};\pmb{\eta}^{n},u_{k}^{n})\triangleq \theta f(\pmb{\eta}^{n},u_{k}^{n})+\frac{2\Pd}{u_k^n}\!\sum\limits _{k'=1}^{K}\!\B{c}^{nT}_{k'}\!\left(\! N^2\bar{\pmb{\gamma}}_{k'k}\bar{\pmb{\gamma}}_{k'k}^{T} + N\pmb{D}_{k'k}^2\right)\!(\dot{\B{c}}_{k'}-\theta\B{c}^{n}_{k'}) -\frac{f(\B{c}^{n},u_{k}^{n})}{u_{k}^{n}}(\dot{u}_{k}-\theta u_{k}^{n}). 
\end{align}
\hrulefill
\end{figure*}
\setcounter{eqncnt}{\value{equation}}
\setcounter{equation}{\value{eqnback}}

Note that \eqref{eq:rotatecone} 
is a rotated cone and admits a SOC representation. We numerically observe that modern convex solvers such as GUROBI \cite{gurobi14} or MOSEK \cite{MOSEK:00:SOFT} can solve \eqref{eq:subprob:cvx} of relatively large size, at least sufficient to characterize the performance of cell-free massive MIMO considered in our paper.  The algorithm for solving \eqref{eq:opt4} can be summarized as follows.

\hrulefill
\begin{algorithm}[SCA algorithm for \eqref{eq:opt4}]\label{sec:algo_SCA} 
\begin{description}
  \item[1.] \emph{Initialization}: set $n=1$, choose the initial point of
  $\left(\dot{\B{c}}, \dot{\B{u}}\right)$ as $\left(\dot{\B{c}}^{1}, \dot{\B{u}}^1 \right)$, where $\dot{\B{u}} \triangleq  [\dot{u}_{1},\ldots,\dot{u}_{K}]^{T}$. Choose the spectral  efficiency targets $\{{\So}_{k}\}$, $k=1, \ldots, K$. Define a tolerance $\epsilon$ and the maximum number of iterations $N_{\text{I}}$.\footnote{It is shown in the numerical results that Algorithm~1 converges  quickly after about $10$ iterations.}
  \item[2.] \emph{Iteration} $n$: solve \eqref{eq:subprob:cvx}.
 Let $\left(\dot{\B{c}}^{\ast}, \dot{\B{t}}^{\ast}, \dot{\B{u}}^{\ast}, \theta^{\ast}\right)$ be the solution.
  \item[3.] If $\left|\sum_{k=1}^K\left(\dot{t}_k^\ast - \dot{t}_k^{(n)}\right) \right| < \epsilon$ or $n=N_{\text{I}}$ $\rightarrow$ Stop. Otherwise, go to step 4.
  \item[4.] Set $n = n+1$, update $\left(\dot{\B{c}}^{n}, \dot{\B{u}}^{n}\right) =\left(\dot{\B{c}}^{\ast},  \dot{\B{u}}^{\ast}\right)$, go
  to step 2.
\end{description}
\end{algorithm}

\hrulefill

\begin{remark}
We  recall that for a given spectral efficiency target ${\So}_{k}$, the feasible set of \eqref{eq:opt1c} is convex, so finding an initial point to start Algorithm \ref{sec:algo_SCA} can be done easily by solving a feasibility SOCP. If the problem is infeasible, we simply set $\Ee=0$.
\end{remark}

\subsection*{Convergence Analysis}
	The convergence analysis of Algorithm \ref{sec:algo_SCA} follows standard arguments for the general framework of SCA \cite{SFL:16:SP,BBT:10:J_GO}. Specifically, the following properties of Algorithm \ref{sec:algo_SCA} are guaranteed
    \begin{itemize}
    \item A feasible solution to  $ (\hat{\mathcal{P}}_{1,n})$ is also feasible to $ (\hat{\mathcal{P}}_{1})$.
    \item An optimal solution of  $ (\hat{\mathcal{P}}_{1,n})$ is also feasible to $ (\hat{\mathcal{P}}_{1,n+1})$.
    \item Algorithm \ref{sec:algo_SCA} generates a monotonically increasing sequence of objectives.
    \end{itemize}   
To see the above results, let us consider the constraint \eqref{eq:SINR:slack:equi}. In iteration $n$, the constraint \eqref{eq:SINR:slack:equi} is replaced by \eqref{eq:rate:1storder}. For ease of description, let $g(\B{c})$ denote the right hand side of the constraint \eqref{eq:SINR:slack:equi}.  Suppose $(\B{c},u_k)$ is feasible to $ (\hat{\mathcal{P}}_{1,n})$, i.e., $F(\B{c},u_k;\B{c}^{n},u_k^{n}) \geq  g(\B{c})$. Since  $ f(\B{c},u_k) $  is jointly convex with $(\B{c},u_k)$, it holds that $ f(\B{c},u_k) \geq F(\B{c},u_k;\B{c}^{n},u_k^{n})$ since $F(\B{c},u_k;\B{c}^{n},u_k^{n})$ is simply the first order approximation of $ f(\B{c},u_k) $ around $(\B{c}^{n},u_k^{n})$,   and thus $ f(\B{c},u_k) \geq g(\B{c})$. This implies that $(\B{c},u_k)$ is also feasible to  $ (\hat{\mathcal{P}}_{1})$. Further, since $(\B{c}^{n},u_k^{n})$  is an optimal solution to    $ (\hat{\mathcal{P}}_{1,n})$, it is of course feasible to      $ (\hat{\mathcal{P}}_{1,n})$ and also to   $ (\hat{\mathcal{P}}_{1})$, i.e.,  $f(\B{c}^{n},u_k^{n}) \geq g(\B{c}^{n})$. Now we note that $F(\B{c}^{n},u_k^{n};\B{c}^{n},u_k^{n})=f(\B{c}^{n},u_k^{n}) $ since $F(\B{c},u_k;\B{c}^{n},u_k^{n})$ is equal to $f(\B{c},u_k)$ when $(\B{c},u_k)=(\B{c}^{n},u_k^{n})$. Thus, $F(\B{c}^{n},u_k^{n};\B{c}^{n},u_k^{n})\geq g(\B{c}^{n})$ which means that  $(\B{c}^{n},u_k^{n})$ is also feasible to $ (\hat{\mathcal{P}}_{1,n+1})$. Obviously, the optimal value of an optimal problem is always larger than the objective value of a feasible solution, which proves the monotonic increase of the objective returned by Algorithm \ref{sec:algo_SCA}.   
We also note that due the total power constraint, the objective of $ ({\mathcal{P}}_{1})$ is bounded from above. Thus, the objective of Algorithm \ref{sec:algo_SCA} is guaranteed to converge.

\section{Access Point Selection}\label{sec:APsel}
Compared with colocated massive MIMO systems, cell-free massive MIMO systems require more backhaul connections to transfer the data between the APs and the CPU. This is reflected by the last  term of \eqref{eq:totalP4} (representing the total power consumption of the backhaul) which is proportional to the sum spectral efficiency and the numbers of APs. By dividing the numerator and the denominator of the total energy efficiency \eqref{eq:EE1} by ${B\cdot\Se}$, we get
\setcounter{equation}{37}\begin{align} \label{eq:APsel1}
\Ee 
	=
\frac{1}{\frac{\bar{P}_{\mathrm{fix}} + \Pd N_0 N\sum\limits_{m=1}^M \frac{1}{\alpha_m}  \sum\limits_{k=1}^K \eta_{mk}\gamma_{mk}}{{B\cdot\Se}}  + \sum\limits_{m=1}^M P_{\text{bt},m}
    }.
\end{align}
We can see that the backhaul power consumption--the second term of the denominator of \eqref{eq:APsel1}--affects significantly the energy efficiency, especially when $M$ increases. To improve the total energy efficiency, we can decrease the first term of the denominator of the energy efficiency in \eqref{eq:APsel1} or/and reduce the backhaul power consumption.  With the proposed power allocation scheme in preceding sections, we just minimize the first term, ignoring the effect of the backhaul power consumption. That is, there is still room to further increase the total energy efficiency of the system.

 In this section, we propose two access point selection schemes which can reduce the backhaul power consumption, and hence, increase the energy efficiency. The proposed schemes are based on two main observations:
\begin{itemize}
\item The backhaul between the CPU and the $m$-th AP is used to transfer the data $q_1, \ldots, q_K$. Thus, the backhaul power consumption depends on the spectral efficiencies ${\Se}_{1}, \ldots, {\Se}_{K}$. If the $m$-th AP serves only some users, then it needs to send only the data corresponding to these users. As a result, the backhaul power consumption depends only on the spectral efficiencies of these users. Let $\U_m$ be the set of users served by the $m$-th AP. Then, the backhaul consumption corresponding to the $m$-th AP \eqref{eq:totalP3} is now modified as
\begin{align} \label{eq:seltotalP3m}
P_{\text{bh},m} = P_{0,m}  +B \cdot \sum_{k\in\U_m}  P_{\text{bt},m}{\Se}_k.
\end{align}
Clearly, if $\U_m =\{1, \ldots, K\}$ for all $m$, then \eqref{eq:seltotalP3m} is identical to \eqref{eq:totalP3}.

\item For a given user, there are many APs which are located very far away. These APs will not add significantly to the overall spatial diversity gains. This implies that not all APs really participate in
serving this user.
\end{itemize}

Motivated by the above observations, and in order to save the power consumption (and therefore, increase the total energy efficiency), each user should not be served by all APs. Instead, a group of APs should be chosen for each user. In this section, we propose two simple AP selection methods: received-power-based selection and largest-large-scale-fading-based selection.

\subsection{Received-Power-Based Selection}\label{subsec:RPBS}
Based on the optimal power control coefficients obtained from Algorithm~\ref{sec:algo_SCA}, we can determine how much useful power is transferred from each AP  to  a  given  user, and hence, we can select a group of APs which effectively serves that user.
With the received-power-based selection scheme, a group of APs   are chosen to serve the $k$-th user, denoted by $\A_k$, which should fulfill the following criteria: (i) it  contributes  at  least  $\delta$\%  of  the total received  power of the desired signal at the $k$-th user, and (ii) its cardinality is minimum.

From \eqref{eq:rateexpr1}, the total received power of the desired signal is represented by ${\tt DS}_k$,
	\begin{align}
	{\tt DS}_k=\sqrt{\Pd}N\sum_{m=1}^M\sqrt{ \eta_{mk}}\gamma_{mk},
	\end{align}
in which the $m$-th AP contributes an amount of $\sqrt{\Pd}N \sqrt{\eta_{mk}}\gamma_{mk}$. Therefore, the $m$-th AP  contributes a fraction
	\begin{align} \label{eq:selperc1}
	p(m,k) =\frac{\sqrt{\eta_{mk}}\gamma_{mk}}{\sum_{m'=1}^M\sqrt{\eta_{m'k}}\gamma_{m'k}}
	\end{align}
 of the total received power at the $k$-th user. As a result, $\A_k$ includes $|\A_k|$ APs that correspond to the $|\A_k|$ largest  $p(m,k)$ and $\sum_{m\in\A_k}p(m,k) \geq \delta\%$. To find 
$\A_k$, we can first order $\{p(m,k)\}$, $m=1, \ldots, M$, in descending order: $p(k^{(1)},k) \leq p(k^{(2)},k)\leq \ldots \leq p(k^{(M)},k)$, where $k^{(m)}\in\{1, \ldots, M\}$. Then
we choose $M_{k}$ so that $\sum_{m=1}^{M_{k}} p(k^{(m)},k) \geq \delta\%$ and $M_{k}$ is minimum. The set $\A_k$ is $\{k^{(1)}, \ldots, k^{(M_k)}\}$.
 After choosing $K$ sets $\A_k$, $k=1, \ldots, K$, we can determine $\U_m$ (recall that $\U_m$ is the set of users served by the $m$-th AP).

With our AP selection scheme, the energy efficiency maximization problem in \eqref{eq:opt1b} becomes
\begin{align} \label{eq:opt1bsel1}
    (\mathcal{P}_2) : \left\{%
\begin{array}{ll}
  \mathop {\max}\limits_{\{\eta_{mk}\}} & \frac{{B\cdot\Se}}{\bar{P}_{\mathrm{fix}} + \Pd N_0 N\sum\limits_{m=1}^M \frac{1}{\alpha_m}  \sum\limits_{k=1}^K \eta_{mk}\gamma_{mk}  + P_{\text{bh,sel}}
    }\\
  \text{s.t.} & {\Se}_{k}\geq {\So}_{k},\\
              & \sum_{k\in\U_m} \eta_{mk}\gamma_{mk} \leq 1/N, ~ \forall m, \\
              & \eta_{mk} \geq 0, ~ \forall k, ~ \forall m,\\
                          & \eta_{mk} = 0, k\notin \U_m, \forall m,\\
\end{array}%
\right.
\end{align}
where $ P_{\text{bh,sel}}\triangleq B \cdot \sum_{m=1}^M\sum_{k\in\U_m}  P_{\text{bt},m}{\Se}_k$.
Since the power control coefficients $\{\eta_{mk}\}$ are always coupled with $\gamma_{mk}$ in the objective function as well as the constraints, the optimization problem $(\mathcal{P}_2)$ is equivalent to 
\begin{align} \label{eq:opt1bsel2}
    (\mathcal{P}_2) : \left\{%
\begin{array}{ll}
  \mathop {\max}\limits_{\{\eta_{mk}\}} & \frac{{B\cdot\Se}}{\bar{P}_{\mathrm{fix}} + \Pd N_0 N\sum\limits_{m=1}^M \frac{1}{\alpha_m}  \sum\limits_{k=1}^K \eta_{mk}\hat{\gamma}_{mk}  + P_{\text{bh,sel}}
    }\\
  \text{s.t.} & {\Se}_{k}\geq {\So}_{k},\\
              & \sum_{k=1}^K \eta_{mk}\hat{\gamma}_{mk} \leq 1/N, ~ \forall m, \\
              & \eta_{mk} \geq 0, ~ \forall k, ~ \forall m,\\
\end{array}%
\right.
\end{align}
where $\hat{\gamma}_{mk} = {\gamma}_{mk}$ when $k\in\U_m$ and $0$ otherwise.

The denominator of the objective function in \eqref{eq:opt1bsel2} involves the discrete sets $\U_m$, $m=1, \ldots, M$. Unfortunately, we  cannot  solve  \eqref{eq:opt1bsel2}  directly  using
convex  optimization  tools. Yet, by using the following bound,
$$\sum_{m=1}^M\sum_{k\in\U_m}  P_{\text{bt},m}{\Se}_k \leq \sum_{m=1}^M\sum_{k=1}^K  P_{\text{bt},m}{\Se}_k = \sum_{m=1}^M  P_{\text{bt},m}{\Se},$$ we can efficiently find an approximate solution of \eqref{eq:opt1bsel2} by replacing $\sum_{m=1}^M\sum_{k\in\U_m}  P_{\text{bt},m}{\Se}_k$ with $\sum_{m=1}^M  P_{\text{bt},m}{\Se}$ which is in turn identical to \eqref{eq:opt1b}, but $\gamma_{mk}$ is now replaced with $\hat{\gamma}_{mk}$.

The algorithm to implement the received-power-based selection scheme is summarized as follows.

\hrulefill
\begin{algorithm}[Received-power-based selection]\label{sec:power_APsel} 
\begin{description}
  \item[1.] Choose $\delta$, perform Algorithm~\ref{sec:algo_SCA} to find optimal power coefficients $\{\eta_{mk}^\ast\}$. Then, compute $p(m,k)$ according to \eqref{eq:selperc1}.
  \item[2.] For each $k$, find set $\A_k$: Sort $p(m,k)$ in descending order $p(k^{(1)},k) \leq p(k^{(2)},k)\leq \ldots \leq p(k^{(M)},k)$, where $k^{(m)}\in\{1, \ldots, M\}$. Let $\A_k =\{k^{(1)}\}$.
  
  \pmb{for} $i=2$ to $M$ \pmb{do}
  
  \quad \pmb{if} ~ $\sum_{m\in\A_k}p(m,k) \geq \delta\%$ ~ \pmb{then} ~ stop,
  
  \quad \pmb{else} ~ $\A_k = \A_k \bigcup \{k^{(i)}\}$,
  
  \quad \pmb{end if}
  
  \pmb{end for}
  \item[3.] From $\A_k$, determine $\U_m$, $m=1, \ldots, M$. Let $\hat{\gamma}_{mk} = {\gamma}_{mk}$ when $k\in\U_m$ and $0$ otherwise, $k=1, \ldots, K$, $m=1, \ldots, M$.

  \item[4.] Use Algorithm~\ref{sec:algo_SCA}, but replace $\gamma_{m,k}$ with $\hat{\gamma}_{mk}$, to find the optimal power coefficients $\{\eta_{mk}\}$.
\end{description}
\end{algorithm}

\hrulefill

\subsection{Largest-Large-Scale-Fading-Based Selection}
In the received-power-based selection scheme, the sets $\A_k$, $k=1, \ldots, K$, are chosen based on the  power control coefficients $\{\eta_{mk}\}$ obtained from Algorithm~\ref{sec:algo_SCA}. It means that we have to perform Algorithm~\ref{sec:algo_SCA} to find $\A_k$ which incurs high computational complexity. In this section, we propose a simpler method, called largest-large-scale-fading-based selection method, which chooses  $\A_k$ without the implementation of Algorithm~\ref{sec:algo_SCA}.  

With the largest-large-scale-fading-based selection method, the $k$-th  user  is  associated  with 
only $M_{0,k} \leq M$ APs corresponding to the $M_{0,k}$ largest large-scale fading coefficients. The main question arising immediately is how to choose $M_{0,k}$. Naturally, we can choose $M_{0,k}$ APs which satisfy 
\begin{align}\label{eq:strong_al1}
\sum_{m=1}^{M_{0,k}}\frac{\bar{\beta}_{mk}}{\sum_{m'=1}^M{\beta_{m'k}}} \geq \delta\%,
\end{align}
where $\{\bar{\beta}_{1k}, \ldots, \bar{\beta}_{Mk}\}$ is the sorted (in descending order) set  of the set $\{\beta_{1k}, \ldots, \beta_{Mk}\}$. After choosing $\A_k$, we can follow the same method as in Section~\ref{subsec:RPBS} to find optimal power control coefficients.

The algorithm to implement the strongest-large-scale-fading-based selection method is summarized as follows.

\hrulefill
\begin{algorithm}[Largest-large-scale-fading-based selection]\label{sec:strongest_APsel} 
\begin{description}
  \item[1.] By using \eqref{eq:strong_al1}, user $k$ chooses a group of $M_{0,k}$ serving APs which correspond to the $M_{0,k}$ largest large-scale fading coefficients. Then, we can determine $\U_m$, $m=1, \ldots, M$.
  \item[2.] Let $\hat{\gamma}_{mk} = {\gamma}_{mk}$ when $k\in\U_m$ and $0$ otherwise, $k=1, \ldots, K$, $m=1, \ldots, M$.

  \item[3.] Use Algorithm~\ref{sec:algo_SCA}, but replace $\gamma_{m,k}$ with $\hat{\gamma}_{mk}$, to find the optimal power coefficients $\{\eta_{mk}\}$.

\end{description}
\end{algorithm}

\hrulefill

\section{Numerical Results and Discussion}\label{sec:numerical_results}
In this section, we provide numerical results to quantitatively study the performance of cell-free massive MIMO in terms of its total energy efficiency, as well as to verify the benefit of our AP selection schemes.

\subsection{Parameters and Setup}
The APs and the users are located within a square of $D\times D$ $\text{km}^2$. The square  is wrapped around at the edges to avoid boundary effects. Furthermore, for simplicity, random pilot assignment is used. With random pilot assignment, each user randomly chooses a pilot sequence from a predefined set of $\taup$ orthogonal pilot sequences of length $\taup$ symbols.
 
The large-scale fading coefficient $\beta_{mk}$  is modeled as the product of path loss and shadow fading:
\begin{align}\label{eq:betmksil}
\beta_{mk} = \text{PL}_{mk}\cdot
z_{mk},
\end{align}
where $z_{mk}$ represents the log-normal shadowing with the standard deviation $\sigma_{\text{sh}}$, and $\text{PL}_{mk}$ represents the three-slope path loss given by (in dB) \cite{TSG:01:VTC}
\begin{align}\label{eq:ploss}
\text{PL}_{mk} \!=\! \left\{\!
\begin{array}{l}
  -L - 35\log_{10} (d_{mk}), ~ \text{if} ~ d_{mk}>d_1\\
  -L - 15\log_{10} (d_1) - 20\log_{10} (d_{mk}), \\ \hspace{3cm}\text{if} ~ d_0< d_{mk}\leq d_1\\
  -L - 15\log_{10} (d_1) - 20\log_{10} (d_{0}), ~ \text{if} ~ d_{mk} \leq d_0\\
\end{array}%
\right.
\end{align}
where $L$ is a constant depending on the carrier frequency, the user and AP heights. In all examples, we choose $\sigma_{\text{sh}}=8$~dB, $d_0=10$~m, $d_1=50$~m, and $L=140.7$~dB. These parameters resemble those in \cite{NAYLM:16:WCOM}.

 Regarding the power consumption parameters, unless otherwise stated, we keep the power consumption parameters as in Table~\ref{table:pcp}. These values are taken from \cite{BSHD:15:WCOM,ZZYJL:16:VT}.  
In addition, we choose $B=20$~MHz, a noise figure equal to 9~dB,  $\Pd=1$~W, $\Pp=0.2$~W, and $\tauc=200$.

\begin{table}[b!]
    \caption{
        Power Consumption parameters.
    }
    \centerline{
\begin{tabular}{|l|l|}
  \hline
  Parameter & Value \\
  \hline
  Power amplifier efficiency, $\alpha_m $, $\forall m$ & $0.4$ \\
    \hline
   Internal power consumption/antenna, $P_{\text{tc},m}$, $\forall m$& $0.2$ W \\
    \hline
  Fixed  power consumption/each  backhaul, $P_{0,m}$, $\forall m$ & $0.825$ W \\
  \hline
 Traffic-dependent backhaul  power, $P_{\text{bt},m}$, $\forall m$ & $0.25$ W/(Gbits/s) \\
  \hline
\end{tabular}}
    \label{table:pcp}
\end{table}

\subsection{Results and Discussions}

\begin{figure}[t!]
\centerline{\includegraphics[width=0.45\textwidth]{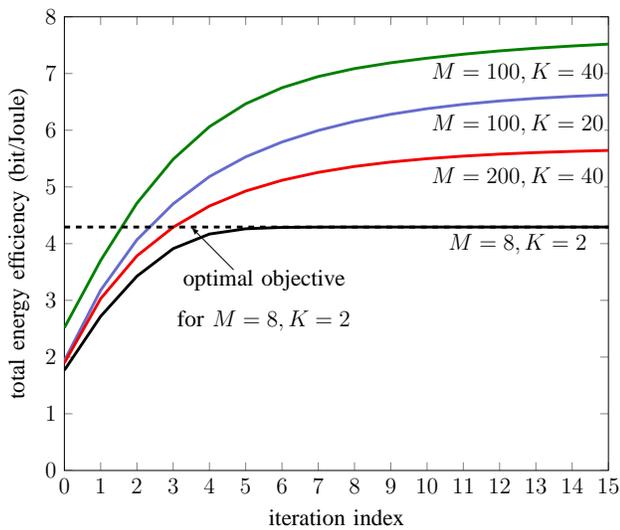}}
\caption{The total energy efficiency versus the number of iterations. Here, 
  $N=1$, and $\taup=20$.
\label{fig:EE_iteration}}
\end{figure}

\subsubsection{Power Allocation}First, we evaluate the effectiveness of our power
allocation in maximizing the total energy efficiency as well as the convergence behavior of Algorithm~\ref{sec:algo_SCA}. Figure~\ref{fig:EE_iteration} shows the total energy efficiency obtained via Algorithm~\ref{sec:algo_SCA} versus the number of iterations $N_{\text{I}}$, with $N=1$, $D=1$~km, $\taup=20$, and different $M$, $K$ for an arbitrary large-scale fading realization. For small network configurations, we benchmark Algorithm~\ref{sec:algo_SCA} with the optimal solution achieved by a BRB method. To do this, we  modify the code for optimal downlink beamforming in \cite{TTJ:16:SP}. Although, there is no guarantee that SCA-based method can yield an optimal solution in theory, our numerical results show that Algorithm~\ref{sec:algo_SCA} indeed converges to an optimal solution for small size networks in most cases. However, the effectiveness of  Algorithm~\ref{sec:algo_SCA} for large-scale networks is still unknown simply because the BRB method fails to work on such large-scale problems. We can also see that  Algorithm~\ref{sec:algo_SCA} converges very fast,
within about $10$ iterations. Therefore, hereafter we choose $N_{\text{I}}=10$
for  Algorithm~\ref{sec:algo_SCA}. Furthermore, we choose $\epsilon=0.01$.

\begin{figure}[t!]
\centerline{\includegraphics[width=0.45\textwidth]{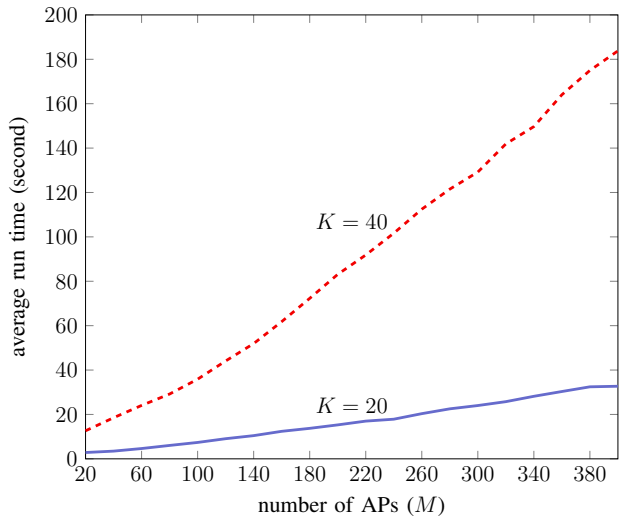}}
\caption{Average computation time (seconds) versus the number of APs. Here, $N=1$, $D=1$~km, and $\taup=20$.
\label{fig:runtime}}
\end{figure}

In the next numerical experiment, we  evaluate the computational complexity of the proposed algorithm. Specifically,  we  provide the average run time of our proposed power control for different $M$ and $K$, as shown in Figure~\ref{fig:runtime}. The codes are developed on MATLAB using the modeling tool YALMIP and are executed on a $64$-bit operating system with $16$~GB RAM and Intel CORE i7, $3.4$~GHz. The convex conic solver MOSEK is chosen as the internal solver for YALMIP. We remark that our codes are not a real-time implementation and, thus, are mainly meaningful to see how the complexity of the proposed algorithm scales with the problem size. As expected, the run time of the proposed algorithms increases with both $M$ and $K$, and increases rapidly after some value of $M$ and $K$. This is a typical observation of interior-point methods for solving convex programs, which are the basis of MOSEK. As the focus of this paper is to investigate the potential of cell-free massive MIMO, we simply resort to off-the-self convex solvers to deal with the derived convex programs. The results shown in Figure~\ref{fig:runtime} clearly call for more computationally efficient methods for optimization problems possibly encountered in cell-free massive MIMO due to their large scale. This opens a rich area for future research.

\begin{figure}[t!]
\centerline{\includegraphics[width=0.45\textwidth]{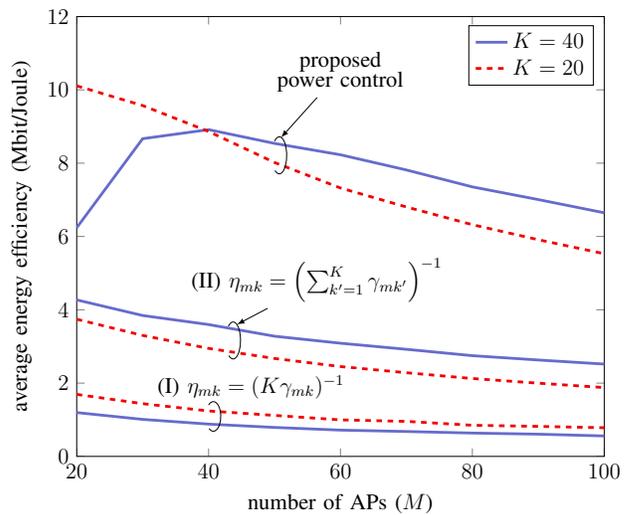}}
\caption{Average total energy efficiency versus the number of APs, with the proposed and equal power control. Here, $N=1$, $D=1$~km, and $\taup=20$.
\label{fig:powerVsnopower}}
\end{figure}

To further evaluate the effectiveness of the proposed power allocation method, we compare the average energy efficiency  of our power allocation with two simple equal power control schemes: (I) $\eta_{mk} = (K\gamma_{mk})^{-1}$ and (II) $\eta_{mk} = \left(\sum_{k'=1}^K\gamma_{mk'} \right)^{-1}$, $\forall k=1, \ldots K$ (see Figure~\ref{fig:powerVsnopower}).\footnote{The power control scheme (I) corresponds to the case where the transmitted powers allocated for all users are the same and equal to $\eta_{mk}\gamma_{mk} = 1/K$. The power control scheme (II) corresponds to the case that all APs use full power, and the power control coefficients $\eta_{mk}$, $k=1,\ldots K$, are the same.} The average is taken over the large-scale fading realizations. The spectral efficiency target ${\So}_{k}$ is set to be equal to the spectral efficiency obtained in scheme (II).  With our proposed power control, the total energy efficiency improves significantly. In particular,  compared with equal power control (II), our proposed
 power control can improve the total energy efficiency by factors of more than $2.9$ and $2.6$ for the cases of ($M=100, K=20$) and ($M=100, K=40$), respectively. Hereafter, all curves are obtained by using power control. Furthermore,   if the problem is infeasible, we set the total energy efficiency equal to $0$.

 \begin{figure}[t!]
\centerline{\includegraphics[width=0.45\textwidth]{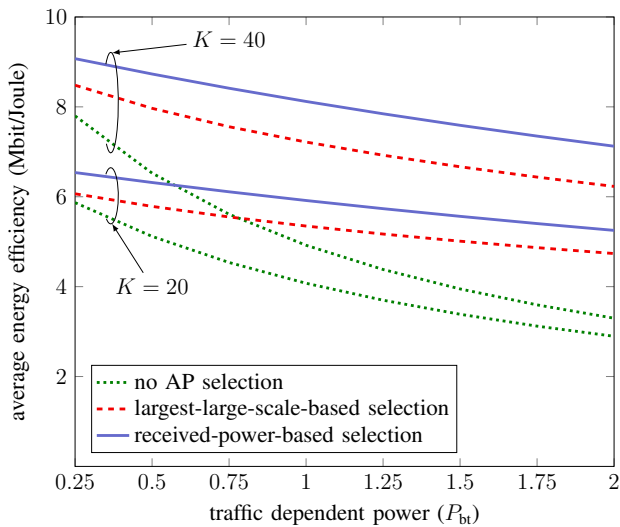}}
\caption{Average total energy efficiency versus the traffic-dependent power, $P_{\text{bt}}$, with and without AP selection schemes. Here, we assume that $P_{\text{bt},m} = P_{\text{bt}}$ for all $m$, $M=100$, $N=1$,  $D=1$~km, and $\taup=40$.
\label{fig:EE_bt}}
\end{figure}

\begin{figure}[t!]
\centerline{\includegraphics[width=0.45\textwidth]{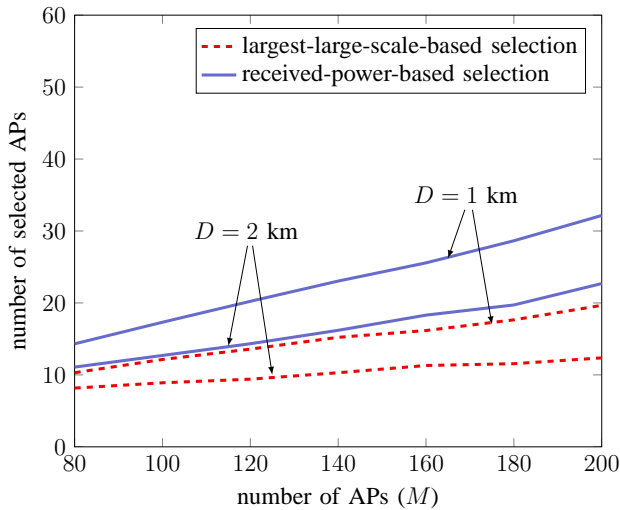}}
\caption{The average number of chosen APs per user. Here,  $N=1$, $K=40$, and $\taup=40$, $S_{\text{o}k} =1$.
\label{fig:selected_AP}}
\end{figure}

\subsubsection{AP Selection} 
Next, we examine the performance of our proposed AP selection schemes and the effect of backhaul power consumption on the energy efficiency. Figure~\ref{fig:EE_bt} shows the average total energy efficiency versus the traffic-dependent  power $P_{\text{bt}}$. Here we assume that $P_{\text{bt},m} = P_{\text{bt}}$ for all $m$, and choose $M=100$, $N=1$, $\taup=40$, $D=1$~km, and the spectral  efficiency target is $1$~bit/s/Hz, i.e., ${\So}_{k}=1$~bit/s/Hz, $k=1, \ldots, K$. For the AP selection schemes, we choose $\delta=95$ which means the chosen APs  contribute  at  least  $95$\%  of  the total received  power of the desired signal at the each user. Without AP selection, the total energy efficiency decreases noticeably as the traffic-dependent power increases. This can be compensated by using our proposed AP selection schemes. Our proposed AP selection schemes work very well and improve the energy efficiency significantly, especially at high $P_{\text{bt},m}$ and large $M$. Furthermore, it can be seen from the figure that, the received-power-based selection scheme is better than the largest large-scale-fading-based selection one. However, the performance gap is small. Note again that the former scheme has higher computational complexity. The average numbers of APs chosen by each user are shown in Figure~\ref{fig:selected_AP}. On average, only about $10\%–-20\%$ of the total number of APs really participate in serving a given user. These results imply that, it is enough for each user to choose a small number of APs around it, which enables the practical implementation of cell-free massive MIMO. In the remaining numerical results, we consider the received-power-based selection scheme with $\delta=95$.

\begin{figure}[t!]
\centerline{\includegraphics[width=0.45\textwidth]{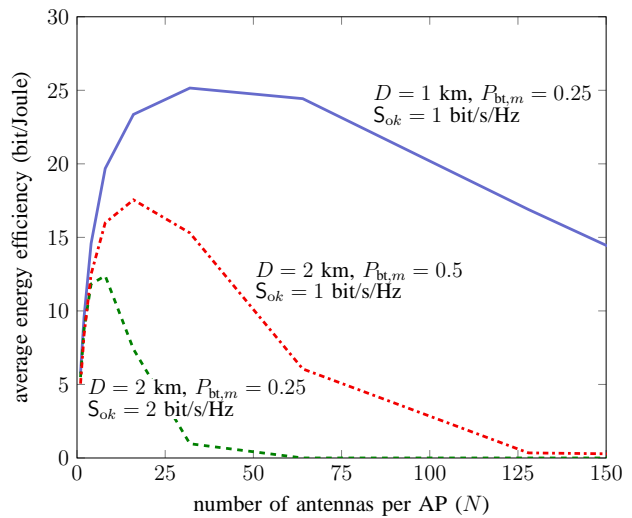}}
\caption{Average total energy efficiency versus the number of  antennas per AP. Here,  $K=40$, and $\taup=40$, and $NM=256$.
\label{fig:EE_N}}
\end{figure}

\begin{figure}[t!]
\centerline{\includegraphics[width=0.45\textwidth]{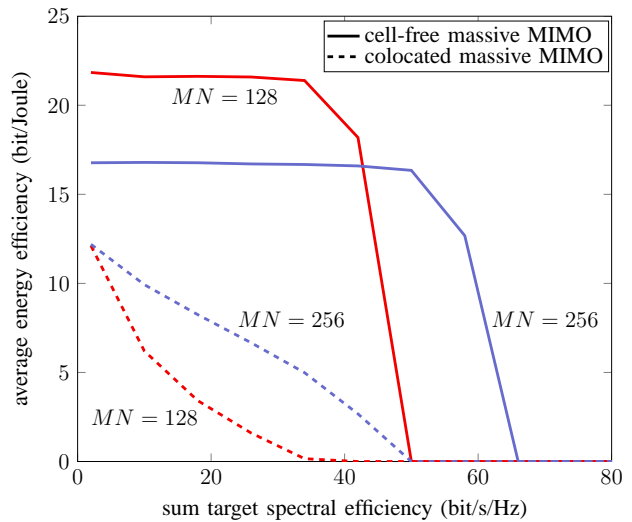}}
\caption{Average total energy efficiency versus sum  spectral efficiency target. Here,  $K=20$, and $\taup=40$, $D=1$~km.
\label{fig:EE_spec}}
\end{figure}

\subsubsection{Effect of the Number of Antennas per AP} 
We examine the performance of cell-free massive MIMO with  different numbers of antennas per AP. For fair comparison, the total transmit power in the network is the same, and the number of service antennas is fixed, i.e., $NM=256$.

Figure~\ref{fig:EE_N} shows the average total energy efficiency as a function of the number of antennas per AP. The curves in this figure are generated by using the received-power-based selection scheme (using Algorithm~\ref{sec:power_APsel}). For a fixed total number of service antennas, when the number of antennas per AP increases, the number of APs reduces. The reduction in the number of APs yields a reduction in  backhaul power consumption. However, at the same time, the spectral efficiency may decrease since the users are now far away from the APs.  This is the reason for why in Figure~\ref{fig:EE_N}, when $N$ increases, the energy efficiency first increases, approaches the optimal point, and then decreases. The optimal number of AP antennas depends heavily on the system parameters, especially the traffic-dependent  power $P_{\text{bt},m}$, the 
spectral  efficiency target ${\So}_{k}$, and the size of the area. As expected, at high $P_{\text{bt},m}$, low ${\So}_{k}$ and $D$, we should use fewer APs with more antennas per AP, and vice versus for low $P_{\text{bt},m}$,  high ${\So}_{k}$ and $D$.

\subsubsection{Cell-Free Massive MIMO Versus Colocated Massive MIMO} 

Finally, we compare the energy efficiency of cell-free massive MIMO and colocated massive MIMO. Colocated massive MIMO corresponds to the case $M=1$. For cell-free massive MIMO, we use the received-power-based selection scheme (Algorithm~\ref{sec:power_APsel}) and the optimal number of antennas per AP (in the sense of maximizing the average energy efficiency). Figure~\ref{fig:EE_spec} shows the energy efficiency versus the sum spectral  efficiency target $\sum_{k=1}^K{\So}_{k}$. Here, we choose $D=1$~km and $K=20$. Compared to the colocated system, cell-free massive MIMO improves the system performance significantly. For example, with ${\So}_{k}=1$~bit/s/Hz, for all $k=1, \ldots, K$, compared with colocated massive MIMO, the energy efficiency of cell-free massive MIMO improves by factors of about $7.4$ and $2.2$ for  $MN=128$ and $MN=256$, respectively. This shows the strong potential of cell-free massive MIMO to boost the energy  and spectral efficiency of 5G networks.

\section{Conclusion}\label{sec:conclusion}

We investigated the total energy efficiency of a cell-free massive MIMO system with multiple-antenna APs, taking into account the effects of channel estimation, power control (which maximizes the total energy efficiency), non-orthogonality of pilot sequences, and   backhaul power consumption. Two AP selection schemes were proposed to reduce the backhaul power consumption. We showed that our power allocation algorithm together with the AP selection schemes can improve the energy efficiency significantly. For a given user, only a small number of APs really participate in serving it.  In particular, the cell-free massive MIMO system can provide a very high energy efficiency, while ensuring uniformly good service for all users. With a requirement of $1$~bit/s/Hz for each user, cell-free massive MIMO can  improve  the energy  efficiency  by an order  of  magnitude, compared to colocated massive MIMO.
 
\appendix

\subsection{Proof of Proposition~\ref{prop1}} \label{app:rate}
From \eqref{eq:rateexpr1}, we need to compute ${\tt DS}_k$,
$\EX{|{\tt BU}_k|^2}$, and $ \EX{|{\tt UI}_{kk'}|^2}$. First, we compute ${\tt DS}_k$. Since   $\tilde{\B{g}}_{mk}$ and $\hat{\B{g}}_{mk}$ are
independent, we have
\begin{align}\label{eq:proof rate1}
    {\tt DS}_k 
    &=  
    \sqrt{\Pd}\EX{\sum_{m=1}^M \sqrt{\eta_{mk}}(\hat{\B{g}}_{mk} + \tilde{\B{g}}_{mk})^T\hat{\B{g}}_{mk}^\ast}\nonumber\\
    &= 
    \sqrt{\Pd}\sum_{m=1}^M \sqrt{\eta_{mk}}\EX{\left\|\hat{\B{g}}_{mk}\right\|^2}=
    \sqrt{\Pd}N\bar{\pmb{\gamma}}_{kk}^{T}\bar{\pmb{\eta}}_{k}.
\end{align}

Next, we compute $\EX{|{\tt BU}_k|^2}$. From \eqref{eq:rat2b}, we have
\begin{align}\label{eq:proof rate2a}
&\EX{|{\tt BU}_k|^2}
	=
    \nonumber\\
  &=
    \Pd\eta_{mk}\sum_{m=1}^M\EX{\left|
  \B{g}_{mk}^T\hat{\B{g}}_{mk}^\ast-
  \EX{  \B{g}_{mk}^T\hat{\B{g}}_{mk}^\ast}\right|^2}\nonumber\\
  &=
  {\Pd}\!\!\sum_{m=1}^M\!
  \eta_{mk}\!\left(\!\EX{\left|\B{g}_{mk}^T\hat{\B{g}}_{mk}^\ast
  \right|^2} - |\EX{\B{g}_{mk}^T\hat{\B{g}}_{mk}^\ast}|^2\!\right).
\end{align}
Using \eqref{eq:proof rate1} and the identity $\EX{\|\hat{\B{g}}_{mk}\|^4} = N(N+1)\gamma_{mk}^2$, we obtain
\begin{align}\label{eq:proof rate2}
&\EX{|{\tt BU}_k|^2}
	\!=\!
  {\Pd}\!\!\sum_{m=1}^M\!\!
  \eta_{mk}\!\left(\EX{\left|\tilde{\B{g}}_{mk}^T\hat{\B{g}}_{mk}^\ast + \|\hat{\B{g}}_{mk}\|^2
  \right|^2} - \gamma_{mk}^2\right)\nonumber\\
  &=
  {\Pd}\sum_{m=1}^M
  \eta_{mk}\left(\EX{\left|\tilde{\B{g}}_{mk}^T\hat{\B{g}}_{mk}^\ast
  \right|^2} + \EX{\|\hat{\B{g}}_{mk}\|^4}- N^2\gamma_{mk}^2\right)
  \nonumber\\
  &=
  \Pd N\!\sum_{m=1}^M\!
  \eta_{mk}\left(\gamma_{mk}(\beta_{mk}-\gamma_{mk}) + (N+1)\gamma_{mk}^2-
  N\gamma_{mk}^2\right)\nonumber\\
  &= {\Pd}N\sum_{m=1}^M
  \eta_{mk} \gamma_{mk}\beta_{mk}.
\end{align}

Finally, we compute $\EX{|{\tt UI}_{kk'}|^2}$.
From \eqref{eq:MMSE est1} and \eqref{eq:rat2c}, we have
\begin{align}\label{eq:proof rate3}
&\EX{|{\tt
UI}_{k'}|^2}=
	\Pd\mathbb{E}\left\{\left|\sum_{m=1}^M
\sqrt{\eta_{mk'}}b_{mk'}\B{g}_{mk}^T \right.\right.\nonumber\\&\hspace{1.7cm}\times		\left.\left. \left(\sqrt{\taup
\Pp}\sum_{k''=1}^K \B{g}_{mk''} \pmb{\varphi}_{k''}^H
\pmb{\varphi}_{k'} + \tilde{\B{w}}_{\p,mk'}
\right)^\ast\right|^2\right\}\nonumber\\
	&=
	\taup\Pp\Pd\EX{\left|\sum_{m=1}^M
\sqrt{\eta_{mk'}}b_{mk'} \|\B{g}_{mk}\|^2\pmb{\varphi}_{k}^H
\pmb{\varphi}_{k'} \right|^2}\nonumber\\
	&
	+
	\taup\Pp\Pd\EX{\left|\sum_{m=1}^M
\sqrt{\eta_{mk'}}b_{mk'}\B{g}_{mk}^T \sum_{k''\neq k}^K \B{g}_{mk''}^\ast \pmb{\varphi}_{k''}^T
\pmb{\varphi}_{k'}^\ast  \right|^2}\nonumber\\&+ 
	\Pd\EX{\left|\sum_{m=1}^M
\sqrt{\eta_{mk'}}b_{mk'}  \B{g}_{mk}^T\tilde{\B{w}}_{\p,mk'}^\ast \right|^2},
\end{align}
where 
\begin{align}\label{eq:proof cmk}
b_{mk} \triangleq
\frac{\sqrt{\taup\Pp}\beta_{mk}}{\taup\Pp\sum_{k'=1}^K\beta_{mk'}\left|\pmb{\varphi}_{k'}^H
\pmb{\varphi}_{k}\right|^2+1}.
\end{align}
Since the elements inside the norm of the second and third terms in \eqref{eq:proof rate3} are uncorrelated, zero mean RVs, we can rewrite \eqref{eq:proof rate3} as
\begin{align}\label{eq:proof rate31}
&\EX{|{\tt
UI}_{k'}|^2}
	\nonumber\\
	&=
	\taup\Pp\Pd\EX{\left|\sum_{m=1}^M
\sqrt{\eta_{mk'}}b_{mk'} \|\B{g}_{mk}\|^2 \right|^2} \left|\pmb{\varphi}_{k}^H
\pmb{\varphi}_{k'}\right|^2\nonumber\\
	&+
	\taup\Pp\Pd\sum_{m=1}^M\sum_{k''\neq k}^K\eta_{mk'}b_{mk'}^2\EX{\left|
\B{g}_{mk}^T  \B{g}_{mk''}^\ast\right|^2}\left| \pmb{\varphi}_{k''}^T
\pmb{\varphi}_{k'}^\ast  \right|^2\nonumber\\&+ 
	\Pd\sum_{m=1}^M\eta_{mk'}b_{mk'}^2\EX{\left|
  \B{g}_{mk}^T\tilde{\B{w}}_{\p,mk'}^\ast \right|^2}\nonumber\\
  &=
	\taup\Pp\Pd\EX{\left|\sum_{m=1}^M
\sqrt{\eta_{mk'}}b_{mk'} \|\B{g}_{mk}\|^2 \right|^2} \left|\pmb{\varphi}_{k}^H
\pmb{\varphi}_{k'}\right|^2\nonumber\\
	&+
	\taup\Pp\Pd N\sum_{m=1}^M\sum_{k''\neq k}^K\eta_{mk'}b_{mk'}^2\beta_{mk}\beta_{mk''}\left| \pmb{\varphi}_{k''}^T
\pmb{\varphi}_{k'}^\ast  \right|^2\nonumber\\
&+ 
	\Pd N\sum_{m=1}^M\eta_{mk'}b_{mk'}^2\beta_{mk}.
\end{align}
The expectation that appears in \eqref{eq:proof rate31} is computed as follows,
\begin{align}\label{eq:proof rate7}
&\EX{\left|\sum_{m=1}^M\!\!
\sqrt{\eta_{mk'}}b_{mk'} \|\B{g}_{mk}\|^2 \right|^2}
	\!=\! \EX{\!\sum_{m=1}^M\!\!
\eta_{mk'}b_{mk'}^2\|\B{g}_{mk}\|^4\!}\nonumber\\
	&+\EX{\sum_{m=1}^M\sum_{n\neq m}^M
\sqrt{\eta_{mk'}\eta_{nk'}}b_{mk'}b_{nk'}\|\B{g}_{mk}\|^2
\|\B{g}_{nk}\|^2}\nonumber\\
&= N(N+1)\sum_{m=1}^M \eta_{mk'}b_{mk'}^2
\beta_{mk}^2\nonumber\\&+ N^2\sum_{m=1}^M\sum_{n\neq m}^M
\sqrt{\eta_{mk'}\eta_{nk'}}b_{mk'}b_{nk'}\beta_{mk}
\beta_{nk}.
\end{align}

Inserting \eqref{eq:proof rate7} into \eqref{eq:proof rate3}, and using \eqref{eq:proof cmk}, we obtain
\begin{align}\label{eq:proof rate9}
\EX{|{\tt UI}_{kk'}|^2}
    &=
    \Pd\left|\pmb{\varphi}_{k}^H \pmb{\varphi}_{k'}\right|^2 N^2
    \left(\sum_{m=1}^M\sqrt{\eta_{mk'}}\gamma_{mk'}\frac{\beta_{mk}}{\beta_{mk'}} \right)^2 \nonumber\\&+ \Pd N \sum_{m=1}^M\eta_{mk'}\gamma_{mk'} \beta_{mk}.
\end{align}

The substitution of \eqref{eq:proof rate1}, \eqref{eq:proof rate2}, and
\eqref{eq:proof rate9} into \eqref{eq:rateexpr1} yields
\eqref{eq:Theo_rateexpr1}.

\subsection{Showing the Equivalence Between  $(\mathcal{P})$ and $(\mathcal{P}_1)$}\label{app_trans1}
From \eqref{eq:EE1}, the optimization problem \eqref{eq:opt1} is
\begin{align} \label{eq:opt1b}
    (\mathcal{P}) : \left\{%
\begin{array}{ll}
  \mathop {\max}\limits_{\{\eta_{mk}\}} & \frac{{B\cdot\Se(\{\eta_{mk}\})}}{P_{\text{total}}
    }\\
  \text{s.t.} & {\Se}_{k}(\{\eta_{mk}\})\geq {\So}_{k},~ \forall k\\
              & \sum_{k=1}^K \eta_{mk}\gamma_{mk} \leq 1/N, ~ \forall m \\
              & \eta_{mk} \geq 0, ~ \forall k, ~ \forall m,\\
\end{array}%
\right.
\end{align}
where $P_{\text{total}}$ is given by \eqref{eq:totalP4}.
By dividing the numerator and the denominator of the objective function of $(\mathcal{P})$ by ${B\cdot \Se(\{\eta_{mk}\})}$, we can see that the objective function  increases as 
$$ \frac{{B\cdot\Se(\{\eta_{mk}\})}}{\bar{P}_{\mathrm{fix}} + \Pd N_0 N\sum\limits_{m=1}^M \frac{1}{\alpha_m } \sum\limits_{k=1}^K \eta_{mk}\gamma_{mk}  
    }.$$
Therefore, the optimization problem $(\mathcal{P})$ is equivalent to $(\mathcal{P}_1)$.


\begin{IEEEbiography}
{Hien Quoc Ngo}  received the B.S. degree in electrical engineering from the Ho Chi Minh City University of Technology, Vietnam, in 2007, the M.S. degree in electronics and radio engineering from Kyung Hee University, South Korea, in 2010, and the Ph.D. degree in communication systems from Link\"oping University (LiU), Sweden, in 2015. In 2014, he visited the Nokia Bell Labs, Murray Hill, New Jersey, USA. From January 2016 to April 2017, Hien Quoc Ngo was a VR researcher at the Department of Electrical Engineering (ISY), LiU. He was also a Visiting Research Fellow at the School of Electronics, Electrical Engineering and Computer Science, Queen's University Belfast, UK, funded by the Swedish Research Council.

Hien Quoc Ngo is currently a Lecturer at Queen's University Belfast, UK. His main research interests include massive (large-scale) MIMO systems, cell-free massive MIMO, physical layer security, and cooperative communications. He has co-authored many research papers in wireless communications and co-authored the Cambridge University Press textbook \emph{Fundamentals of Massive MIMO} (2016).

Dr. Hien Quoc Ngo received the IEEE ComSoc Stephen O. Rice Prize in Communications Theory in 2015 and the IEEE Communications Society Leonard G. Abraham Prize in 2017. He also received the IEEE Sweden VT-COM-IT Joint Chapter Best Student Journal Paper Award in 2015. He was an \emph{IEEE Communications Letters} exemplary reviewer for 2014, an \emph{IEEE Transactions on Communications} exemplary reviewer for 2015, and an \emph{IEEE Wireless Communications Letters} exemplary reviewer for 2016. He is  an Editor of Digital Signal Processing. He was a Guest Editor of IET Communications, special issue on ``Recent Advances on 5G Communications'' and a Guest Editor of  IEEE Access, special issue on ``Modelling, Analysis, and Design of 5G Ultra-Dense Networks'', in 2017. He has been a member of Technical Program Committees for several IEEE conferences such as ICC, Globecom, WCNC, VTC, WCSP. 
\end{IEEEbiography}

\begin{IEEEbiography}
{Le-Nam Tran}
 (M'10--SM'17) received the B.S. degree in electrical engineering from Ho Chi Minh City University of Technology, Ho Chi Minh City, Vietnam, in 2003 and the M.S. and Ph.D. degrees in radio engineering from Kyung Hee University, Seoul, Korea, in 2006 and 2009, respectively. He is currently a Lecturer/Assistant Professor at the School of Electrical and Electronic Engineering, University College Dublin, Ireland. Prior to this, he was a Lecturer at the Department of Electronic Engineering, Maynooth University, Co. Kildare, Ireland. From 2010 to 2014, he had held postdoc positions at the Signal Processing Laboratory, ACCESS Linnaeus Centre, KTH Royal Institute of Technology, Stockholm, Sweden (2010-2011), and at Centre for Wireless Communications and the Department of Communications Engineering, University of Oulu, Finland (2011-2014). His  research interests are mainly on applications of optimization techniques on wireless communications design. Some recent topics include energy-efficient communications,  cloud radio access networks,  massive MIMO, and full-duplex transmission. He has authored or co-authored in some 70 papers published in international journals and conference proceedings. 
 
 Dr. Tran is an Associate Editor of  EURASIP Journal on Wireless Communications and Networking. He was Symposium Co-Chair of Cognitive Computing and Networking Symposium of International Conference on Computing, Networking and Communication (ICNC 2016).
\end{IEEEbiography}

\begin{IEEEbiography}
{Trung Q. Duong}(S'05, M'12, SM'13) received his Ph.D. degree in Telecommunications Systems from Blekinge Institute of Technology (BTH), Sweden in 2012. Since 2013, he has joined Queen's University Belfast, UK as a Lecturer (Assistant Professor). His current research interests include small-cell networks, ultra-dense networks, physical layer security, energy-harvesting communications, massive MIMO. He is the author or co-author of more than 270 technical papers published in scientific journals (145 articles) and presented at international conferences (125 papers).

Dr. Duong currently serves as an Editor for the \textsc{IEEE Transactions on Wireless Communications}, \textsc{IEEE  Transactions on Communications}, \textsc{IET Communications}, and a Senior Editor for \textsc{IEEE Communications Letters}. He was awarded the Best Paper Award at the IEEE Vehicular Technology Conference (VTC-Spring) in 2013, IEEE International Conference on Communications (ICC) 2014, and IEEE Global Communications Conference (GLOBECOM) 2016. He is the recipient of prestigious Royal Academy of Engineering Research Fellowship (2016-2021). 
\end{IEEEbiography}

\begin{IEEEbiography}
{Michail Matthaiou}(S'05--M'08--SM'13) was born in Thessaloniki, Greece in 1981. He obtained the Diploma degree (5 years) in Electrical and Computer Engineering from the Aristotle University of Thessaloniki, Greece in 2004. He then received the M.Sc. (with distinction) in Communication Systems and Signal Processing from the University of Bristol, U.K. and Ph.D. degrees from the University of Edinburgh, U.K. in 2005 and 2008, respectively. From September 2008 through May 2010, he was with the Institute for Circuit Theory and Signal Processing, Munich University of Technology (TUM), Germany working as a Postdoctoral Research Associate. He is currently a Reader (equivalent to Associate Professor) in Multiple-Antenna Systems at Queen's University Belfast, U.K. after holding an Assistant Professor position at Chalmers University of Technology, Sweden. His research interests span signal processing for wireless communications, massive MIMO, hardware-constrained communications, and performance analysis of fading channels.

Dr. Matthaiou will receive the 2017 IEEE Communications Society Leonard G. Abraham Prize. He was the recipient of the 2011 IEEE ComSoc Best Young Researcher Award for the Europe, Middle East and Africa Region and a co-recipient of the 2006 IEEE Communications Chapter Project Prize for the best M.Sc. dissertation in the area of communications. He was co-recipient of the Best Paper Award at the 2014 IEEE International Conference on Communications (ICC) and was an Exemplary Reviewer for \textsc{IEEE Communications Letters} for 2010. In 2014, he received the Research Fund for International Young Scientists from the National Natural Science Foundation of China. In the past, he was an Associate Editor for the \textsc{IEEE Transactions on Communications}, Associate Editor/Senior Editor for \textsc{IEEE Communications Letters} and was the Lead Guest Editor of the special issue on ``Large-scale multiple antenna wireless systems'' of the \textsc{IEEE Journal on Selected Areas in Communications}. He was the chair of the Wireless Communications Symposium (WCS) at IEEE GLOBECOM 2016. 
\end{IEEEbiography}

\begin{IEEEbiography}
{Erik G. Larsson} received the Ph.D. degree from Uppsala University,
Uppsala, Sweden, in 2002.

He is currently Professor of Communication Systems at Link\"oping
University (LiU) in Link\"oping, Sweden. He was with the Royal
Institute of Technology (KTH) in Stockholm, Sweden, the University of
Florida, USA, the George Washington University, USA, and Ericsson
Research, Sweden.  In 2015 he was a Visiting Fellow at Princeton
University, USA, for four months.  His main professional interests are
within the areas of wireless communications and signal processing. He
has co-authored some 130 journal papers on these topics, he is
co-author of the two Cambridge University Press textbooks
\emph{Space-Time Block Coding for Wireless Communications} (2003) and
\emph{Fundamentals of Massive MIMO} (2016). He is co-inventor on 16
issued and many pending patents on wireless technology.

He was Associate Editor for, among others, the \emph{IEEE Transactions
  on Communications} (2010-2014) and the \emph{IEEE Transactions on
  Signal Processing} (2006-2010).  From 2015 to 2016 he served as
chair of the IEEE Signal Processing Society SPCOM technical committee,
and in 2017 he is the past chair of this committee.  From 2014 to 2015
he served as chair of the steering committee for the \emph{IEEE
  Wireless Communications Letters}.  He was the General Chair of the
Asilomar Conference on Signals, Systems and Computers in 2015, and its
Technical Chair in 2012.  He is a member of the IEEE Signal Processing
Society Awards Board during 2017--2019.

He received the IEEE Signal Processing Magazine Best Column Award
twice, in 2012 and 2014, the IEEE ComSoc Stephen O. Rice Prize in
Communications Theory in 2015, and the IEEE ComSoc Leonard G. Abraham
Prize in 2017. He is a Fellow of the IEEE.

\end{IEEEbiography}

\end{document}